\documentclass[11pt, a4paper]{article}
\usepackage{jheppub}
\usepackage{amsfonts}
\usepackage{amsthm}
\usepackage{braket}
\usepackage{graphicx}
\usepackage{color}
\usepackage{dsfont}
\usepackage{verbatim}
\usepackage{amscd}
\usepackage{multirow}
\usepackage{subfigure}
\usepackage{tikz}
\usepackage{tikz-feynman}
\tikzfeynmanset{compat=1.0.0}

\usepackage{commath} 
\usepackage{slashed}
\usepackage{braket}

\newcommand{\be}{\begin{equation}}
\newcommand{\ee}{\end{equation}}
\newcommand{\bea}{\begin{eqnarray}}
\newcommand{\eea}{\end{eqnarray}}

\newcommand{\diagramw }[2]{\parbox{#1}{\includegraphics[width=#1]{#2}}}
\newcommand{\diagramwt}[2]{\parbox{#1}{\includegraphics[trim = 5mm 0mm 0mm 0mm, clip, width=#1]{#2}}  }

\begin{document}

\title{Lattice regularisation and entanglement structure of the Gross-Neveu model}

\author[a]{Gertian Roose,}
\author[a,b]{Nick Bultinck,}
\author[a]{Laurens Vanderstraeten,}
\author[a]{Frank Verstraete,}
\author[a]{Karel Van Acoleyen}
\author[a]{and Jutho Haegeman}

\affiliation[a]{Department of Physics and Astronomy, University of Ghent, Krijgslaan 281, 9000 Gent, Belgium}
\affiliation[b]{Department of Physics, University of California, Berkeley, CA 94720, USA}

\emailAdd{gertian.roose@ugent.be}
\emailAdd{nibultinck@gmail.com}
\emailAdd{laurens.vanderstraeten@ugent.be}
\emailAdd{frank.verstraete@ugent.be}
\emailAdd{karel.vanacoleyen@ugent.be}
\emailAdd{jutho.haegeman@ugent.be}


\arxivnumber{2010.03441}

\abstract{
	We construct a Hamiltonian lattice regularisation of the $N$-flavour Gross-Neveu model that manifestly respects the full $\mathsf{O}(2N)$ symmetry, preventing the appearance of any unwanted marginal perturbations to the quantum field theory. In the context of this lattice model, the dynamical mass generation is intimately related to the Coleman-Mermin-Wagner and Lieb-Schultz-Mattis theorems. In particular, the model can be interpreted as lying at the first order phase transition line between a trivial and symmetry-protected topological (SPT) phase, which explains the degeneracy of the elementary kink excitations. We show that our Hamiltonian model can be solved analytically in the large $N$ limit, producing the correct expression for the mass gap. Furthermore, we perform extensive numerical matrix product state simulations for $N=2$, thereby recovering the emergent Lorentz symmetry and the proper non-perturbative mass gap scaling in the continuum limit. Finally, our simulations also reveal how the continuum limit manifests itself in the entanglement spectrum. As expected from conformal field theory we find two conformal towers, one tower spanned by the linear representations of $\mathsf{O}(4)$, corresponding to the trivial phase, and the other by the projective (\textit{i.e.}~spinor) representations, corresponding to the SPT phase.    
}

\maketitle	
\flushbottom

\section{Introduction}
\label{sec:intro}

Lattice field theory, and in particular, lattice gauge theory, has been among the most successful techniques to probe the non-perturbative behaviour of quantum field theories (QFTs), such as those appearing in the standard model. The accurate determination of the proton and neutron masses has been one of the most noteworthy triumphs resulting from this effort. The default approach is to apply Monte Carlo sampling to the path integral in discretised Euclidean spacetime \cite{aubin2004light,davies2004high,kronfeld2012twenty,aoki2014review,aoki2020flag}.

In recent years, the use of tensor network methods has been proposed as an alternative \cite{banuls2020simulating}, with the promise that these are able to access dynamical information and do not suffer from sign problems in the case of fermionic densities or far-from-equilibrium situations \cite{fodor2002new}. One can apply tensor renormalisation group techniques as an alternative to Monte Carlo sampling to the path integral in discretised spacetime \cite{shimizu2012tensor,shimizu2012analysis,liu2013exact,shimizu2014grassmann,shimizu2014critical,takeda2015grassmann,shimizu2018berezinskii,campos2019tensor,butt2020tensor,vanhecke2019scaling,delcamp2020computing}. Alternatively, one can target the wave functional using a tensor network ansatz and apply variational techniques using the field theory hamiltonian (where only the spatial dimensions are discretised) \cite{byrnes2002density,sugihara2004density,sugihara2005matrix,weir2010studying,banuls2013mass,milsted2013matrix,buyens2014matrix,tagliacozzo2014tensor,silvi2014lattice,haegeman2015gauging,kuhn2015non,banuls2015thermal,pichler2016real,milsted2016matrix,buyens2016confinement,zohar2016building,banuls2016chiral,buyens2016hamiltonian,buyens2017real,silvi2017finite,buyens2017finite,bruckmann20193,silvi2019tensor,banuls2019phase,10.21468/SciPostPhysLectNotes.12,funcke2020topological}. This approach is also closely related to the various experiments and proposals for the analog or digital quantum simulation of lattice field theory using various platforms such as trapped ions, superconducting circuits or cold atoms in optical lattices (see Ref.~\cite{banuls2020simulating} and references therein). Aside from preliminary explorations of $\mathbb{Z}_2$ and $\mathsf{U}(1)$ gauge theories in (2+1) dimensions \cite{sugihara2005matrix,tagliacozzo2014tensor,haegeman2015gauging,zohar2016building,10.21468/SciPostPhysLectNotes.12}, most of the tensor network effort has so far been invested in QFTs in (1+1) dimensions, and in particular the $\lambda \phi^4$ model \cite{sugihara2004density,weir2010studying,shimizu2012tensor,shimizu2012analysis,milsted2013matrix,vanhecke2019scaling,delcamp2020computing} and the Schwinger model, i.e.\ (1+1)-dimensional quantum electrodynamics, (as well as non-abelian generalizations thereof)\cite{byrnes2002density,banuls2013mass,buyens2014matrix,silvi2014lattice,kuhn2015non,banuls2015thermal,pichler2016real,buyens2016confinement,banuls2016chiral,buyens2016hamiltonian,buyens2017real,silvi2017finite,buyens2017finite,silvi2019tensor,funcke2020topological}. These models are superrenormalizable, meaning that the coupling constant has a positive mass dimension and sets the energy scale. The relation between the lattice and continuum parameters is governed by a limited number of divergent diagrams, and observables converge like power laws in the lattice spacing $a$ as the continuum limit is approached.

In this manuscript, we use lattice field theory and tensor network tools (numerical and analytical) to probe the non-perturbative properties of the Gross-Neveu (GN) model \cite{gross1974dynamical}, a (1+1)-dimensional model of $N$ massless but interacting fermion flavours, which shares several non-perturbative features with (3+1)-dimensional quantum chromodynamics (see Refs.~\cite{takeda2015grassmann,banuls2019phase} for tensor network studies of the closely related Thirring model, an integrable model for a single massive interacting fermion flavour). The GN interaction has a discrete chiral symmetry and is marginally relevant, (i.e.\ renormalisable and asymptotically free). The interaction term leads to spontaneous breaking of this chiral symmetry and, associated with this, dynamical mass generation. Here observables converge logarithmically slow as the continuum limit is reached. This increases the importance of symmetries prohibiting the presence of other marginally relevant perturbations that could spoil the already slow convergence. Indeed, it turns out to be crucial to meticulously construct the lattice Hamiltonian so as to maximally preserve the symmetries of the field theory, in order to reliably obtain the continuum limit.

Being a paradigmatic model, the GN model has been the subject of several numerical and theoretical studies. Theoretical studies have focused on determining the scattering matrix and full excitation spectrum \cite{zamolodchikov1978exact,witten1978some,karowski1981complete}, as well as a precise determination of the mass gap \cite{forgacs1991exact,forgacs1991exactb,verschelde1997non,van2002dynamical,kneur2002borel,kneur2010renormalization} using a variety of techniques, including thermodynamic Bethe ansatz, large $N$ expansions and the variationally optimised renormalisation group. Most numerical lattice studies use Monte Carlo techniques on the Euclidean lattice, where the fermions are dealt with by replacing the four point interaction by a coupling to an auxiliary bosonic field (using a Hubbard-Stratonovich transformation) and integrating out the resulting quadratic fermion terms, leaving the calculation of the fermion determinant as a computational problem \cite{cohen1983monte,campostrini1989gross,bietenholz1995perfect}. In particular, there has been interest in the phase diagram at finite temperature and chemical potential, and the possible existence of an inhomogeneous phase \cite{karsch1987gross,lenz2020inhomogeneous}.

A lattice prescription of the kinetic term of the fermion model can be obtained using the Wilson prescription \cite{wilson1977quarks} or using the staggered formulation of Kogut and Susskind \cite{kogut1975hamiltonian,susskind1977lattice}. The former explicitly breaks the chiral symmetry, resulting in additive mass corrections that need to be compensated by a properly tuned bare mass term, in order to reach the continuum limit. Furthermore, the Wilson prescription also leads to Aoki phases \cite{aoki1984new,aoki1986recovery}, where reflection (parity) symmetry is broken and a pseudoscalar condensate is formed. Triggered by interest from the optical lattice community, the phase diagrams of this `Gross-Neveu-Wilson' lattice model and its chiral extension in the limits $N\to\infty$ and $N=1$ were studied in recent publications \cite{bermudez2018gross,kuno2019phase}, and feature both trivial, topological and symmetry broken Aoki phases.

The staggered formulation, on the other hand, exhibits remnant lattice symmetries which prohibit perturbative mass corrections. For the particular case of a lattice Hamiltonian (i.e.\ continuous time) in (1+1) dimension, this remnant symmetry corresponds to full translation invariance of the staggered model \cite{susskind1977lattice}. Spontaneous breaking of discrete chiral symmetry can then be related to Peierls dimerisation, so that the GN model with $N=2$ also arises as a continuum description of polyacetylene. In particular, the GN model provides a good description of the resulting topological soliton (kink) that interpolates between the two ground states \cite{campbell1982soliton,chodos1994gross} and which is traditionally described as an explicit domain wall in the Su-Schrieffer-Heeger (SSH) model \cite{su1980solitons}. While the GN model as low-energy field theory in the context of the SSH model is well documented \cite{fradkin1983phase}, the reverse direction where the SSH model is used as inspiration to construct a precise lattice regularisation of the GN model was, to the best of our knowledge, not considered.

The outline of this paper is as follows. Section~\ref{sec:GN_nut} summarises the field theoretic description of the model. In Section~\ref{sec:latham} we construct the lattice model and discuss the dynamical mass generation and kink degeneracy from a condensed matter perspective. In section~\ref{sec:meanfield}, a large-$N$ mean-field solution is given and found to be consistent with the large $N$ field theory. Section~\ref{sec:mps} introduces the symmetric uniform matrix product state (MPS) ansatz, which is then used in section~\ref{sec:MPSgap} to numerically probe the low-energy behaviour of the model for $N=2$. In section~\ref{sec:entanglement}, we discuss the continuum limit from the point of view of entanglement. Finally, section~\ref{sec:ending} provides a concluding discussion and outlook.

\section{Gross-Neveu model in a nutshell}
\label{sec:GN_nut}
We first provide a short introduction to the GN model and its symmetries before porting it to the lattice. The Lagrangian density for the GN field theory reads
\begin{align}
\mathcal{L} = \sum_{c = 1}^{N} \bar{\psi}_c \mathrm{i}\slashed\partial \psi_c + \frac{g^2}{2}\del{\sum_{c=1}^N \bar{\psi}_c\psi_c }^2
\label{GN:QFT}
\end{align}
where $c$ labels the $N$ different flavours or colours of fermions, $\psi_c$ is the two-component Dirac spinor for flavour $c$, $\slashed\partial = \gamma^0 \partial_0 + \gamma^1\partial_1$ and $\bar{\psi}_c = \psi^\dagger_c \gamma^0$, where $\{\gamma^\mu,\gamma^\nu\} = 2g^{\mu\nu}$ and $g^{\mu\nu}$ is the inverse metric tensor. 

The model has an obvious $\mathsf{SU}(N)$ flavour mixing symmetry that can be extended to an $\mathsf{O}(2N)$ symmetry, which also includes the total $\mathsf{U}(1)$ particle number symmetry and the charge conjugation symmetry (which relates the two disconnected components of $\mathsf{O}(2N)$). This $\mathsf{O}(2N)$ symmetry is made explicit by rewriting the Dirac spinor in terms of its Majorana components. By choosing a specific set of gamma matrices where both $\gamma^0$ and $\gamma^1$ are strictly imaginary (so that $ \beta = \gamma^0$ is imaginary and thus antisymmetric, whereas $\alpha = \gamma^0 \gamma^1$ is real symmetric), the Majorana components correspond to the real and imaginary components of the Dirac spinor, i.e. $\psi_c = (\lambda_{2c-1} + \mathrm{i}\lambda_{2c})/\sqrt{2}$ and thus $\lambda_{2c-1} = (\psi_c + \psi_c^\ast)/\sqrt{2}$ and $\lambda_{2c} = -\mathrm{i} \del{\psi_c - \psi_c^\ast}/\sqrt{2}$, which then yields
\begin{align}
\mathcal{L} = \sum_{m=1}^{2N}\overline{\lambda}_m \mathrm{i}\slashed\partial \lambda_m + \frac{g^2}{2}\del{\sum_{m=1}^{2N}\overline{\lambda}_m\lambda_m }^2
\end{align}
with $\bar{\lambda}_m = \lambda_m^{\mathrm{T}} \gamma^0$. This formulation shows the explicit invariance under $\lambda_m \to O_{mn} \lambda_n$ for $O \in \mathsf{O}(2N)$. Coleman's theorem for relativistic theories \cite{coleman1973goldstone}, related to the Mermin-Wagner theorem in condensed matter or statistical physics \cite{MerminWagner}, guarantees that this continuous symmetry cannot be broken and is thus present in the spectrum of the theory.

The GN model has an additional  $\mathbb{Z}_2$ chiral symmetry that acts as $\psi \rightarrow \gamma_5 \psi$ and prohibits perturbative contributions to the condensate $\sigma = \sum_{c\in N}\braket{\bar{\psi}_c\psi_c}$, or thus, a perturbative mass term. Nonetheless, this $\mathbb{Z}_2$ symmetry is spontaneously broken and gives rise to a non-perturbative mass scale. The effect that the ground state exhibits a dimensionful condensate, despite the absence of dimensionful parameters, other than the ultraviolet (UV) regulator scale, is known as dimensional transmutation. The required renormalization group (RG) invariant mass scale can be obtained from the $\beta$\;function 
\begin{align}
	\beta(g) = \frac{\mathrm{d} g}{\mathrm{d} \log \mu} = \beta_0 g^3 + \beta_1 g^5 + \mathcal{O}(g^7)
	\label{beta}
\end{align}
as 
\begin{align}
\Lambda &= \mu \text{e}^{-\int^{g(\mu)} \beta(g)^{-1} \text{d} g} =  \mu \del{-\beta_0 g^2}^{\frac{\beta_1}{2\beta_0^2}} e^{\frac{1}{2\beta_0g^2}} \left[ 1  +       \mathcal{O}(g^2)\right] \label{RGmass}
\end{align}
where $\mu$ is the regulator scale. This mass scale $\Lambda$ is only an infrared (IR) scale for asymptotically free theories ($\beta_0 < 0$). Furthermore, the mechanism by which it enters the IR theory (if any) must necessarily be non-perturbative. For the $\mathsf{O}(2N)$-symmetric GN model, the first terms in the $\beta$\;function were calculated to be $\beta_0=-\frac{N-1}{2\pi}$  and $\beta_1 = \frac{N-1}{4\pi^2}$ \cite{wetzel1984two,gracey1991computation}.

The condensation of $\sigma$ gives rise to a rich spectrum of massive particles \cite{dashen1975semiclassical,zamolodchikov1978exact,karowski1981complete, Fendley_saleur}. Given the $\mathsf{O}(2N)$ symmetry, some understanding of the representation theory of the corresponding Lie-algebra $\mathfrak{so}(2N)$ is useful to label this spectrum. There are two distinct fundamental (half) spin representations of dimension $2^{N-1}$, which are transformed into each other by conjugation or by application of a reflection element from $\mathsf{O}(2N)$ (determinant $-1$). The other fundamental representations $r=1,\ldots,N-2$ are tensor representations, with $r=1$ the defining (vector) representation. We also refer to the spinor representations as projective representations, which generalise the concept of `representations up to a phase' ---as opposed to linear representations such as the tensor representations--- to arbitrary groups.

The spectrum of the GN model contains both trivial and topological excitations, \textit{i.e.}\ kinks that interpolate between the two vacuum states.  Unlike in conventional (\textit{i.e.}\ Ising-type) $\mathbb{Z}_2$ symmetry breaking, where the kink from one vacuum to the other is unique, in the case of GN the kinks are of the Callen-Coleman-Gross-Zee type \cite{feinberg1995kinks} and bind massless fermions. They transform according to the fundamental spinor representations \cite{witten1978some}. This is similar to Jackiw-Rebbi kinks \cite{jackiw1976solitons} and we will interpret this from a condensed matter perspective as the protected gapless edge modes on the interface between a trivial and SPT phase, when constructing the lattice model. Trivial elementary excitations are labelled by a principal quantum number $n=1,\ldots,N-2$ and have a mass $m_n$ relative to the kink mass $m_K$ given by \cite{dashen1975semiclassical,karowski1981complete}
\begin{align}
m_n = 2 m_K \sin \frac{\pi n}{2N-2}
\end{align}
For every $n$, there are multiplets of these excitations labeled by $r = n, n-2, \ldots, \geq 0$, the linear fundamental representations of $\mathfrak{so}(2N)$. These multiplets are fermionic (bosonic) for $r$ (and thus also $n$) odd (even). The elementary fermion corresponds to $n=1$ and thus transforms according to the defining vector representation of $\mathsf{O}(2N)$. An exact result for the mass of this elementary fermion was derived in Ref.~\cite{forgacs1991exact}, namely
\begin{equation}
m_{1} = \frac{\del{4e}^\frac{1}{2N-2}}{\Gamma\del{1-\frac{1}{2N-2}}}\Lambda_{\overline{\text{MS}}}\label{eq:exactmassgap}
\end{equation}
in terms of a specific choice for the RG-invariant scale $\Lambda_{\overline{\text{MS}}}$, known as the modified minimal subtraction scheme when using dimensional regularisation. Note that for $N=2$, the elementary fermion is not stable and decays into two kinks, i.e.\ $m_1 = 2 m_K$. In that case, Eq.~\eqref{eq:exactmassgap} is providing a definition for (twice) the kink mass $m_K$.

When using a different regularisation scheme, such as the lattice Hamiltonian introduced next, the coupling and its UV dependence differ. As a result, the RG independent scales $\Lambda$ defined from Eq.~\eqref{RGmass} need to be matched between different regularisation schemes. In what follows, we obtain $\Lambda_{\overline{\text{MS}}}=\frac{8}{e}\Lambda_{\text{lat}}$ using an exact solution of the lattice model in the limit $N \to \infty$. A more standard yet involved Feynman diagram calculation of the scattering matrix in Appendix~A proves that this relation is valid for all values of $N$.

\section{Lattice Hamiltonian}
\label{sec:latham}

To construct a lattice regulated version of the GN Hamiltonian, we follow the staggered fermion formulation from Ref.~\cite{susskind1977lattice}. While this procedure is well known, we review it with some detail, in order to properly motivate our lattice proposal for the GN interaction.

One interpretation of the staggering procedure, which is useful for what follows, is to discretise the two components of the Dirac spinor at positions differing by half a lattice spacing, i.e.\ $\psi_{c,1}(n a) \to \phi_{c,2n}/\sqrt{a}$ and $\psi_{c,2}((n+\frac{1}{2})a) \to \phi_{c,2n+1}/\sqrt{a}$, with $a$ the lattice spacing. Furthermore, we choose the matrix $\alpha = \gamma^0 \gamma^1$ appearing in the kinetic term of the Dirac Hamiltonian off-diagonal. With this choice the free massless Dirac Hamiltonian only couples derivatives of the first component to the second component of the Dirac spinor, and vice versa. For such terms a symmetric finite difference approximation of the derivative \footnote{The staggered fermion formulation is often introduced with the two spinor components discretized at the same positions and an asymmetric finite difference scheme, e.g. forward for $\partial\psi_1$ and backward for $\partial\psi_2$. Explicitly shifting the discretisation positions of the two components leads to the same end result, but is somewhat more aesthetically pleasing, and also clarifies how to deal with terms where both components (not their derivates) appear on the same position, e.g. $\psi_1^\dagger \psi_2$.} leads to e.g.
\begin{align}
\int \psi_{c,2}^\dagger(x) \partial_x \psi_{c,1}(x) dx &\to \sum_n a \phantom{a}\psi_{c,2}^\dagger(n+\frac{1}{2})a) \partial_x \psi_{c,1}((n+\frac{1}{2})a) \nonumber \\ 
&\to \sum_n a\phantom{a} \psi_{c,2}^\dagger((n+\frac{1}{2}) a) \frac{\psi_{c,1}((n+1)a) - \psi_{c,1}(na)}{a} \nonumber\\
&\to \sum_n \frac{1}{a}\phantom{a}\phi^\dagger_{c,2n+1} \del{\phi_{c,2n+2} - \phi_{c,2n}}
\end{align}
Combined with the requirement that $\alpha$ is real to make the $\mathsf{O}(2N)$ symmetry explicit (and thus easier to preserve in the lattice model) leads to $\alpha = \sigma^x$, and the resulting lattice Hamiltonian is given by
\begin{equation}
aH=\sum_{n} K_{n,n+1}\label{eq:dirackinetic}
\end{equation}
with $K$ the ($N$-flavor) tight-binding or hopping operator
\begin{align}
	K_{n,n+1} &=  \sum_{c =1}^{N} (-\mathrm{i})(\phi_{c,n}^\dagger \phi_{c,n+1} - \phi_{c,n+1}^\dagger \phi_{c,n})= \sum_{m = 1}^{2N} (-\mathrm{i}) \chi_{m,n} \chi_{m,n+1}  \label{eq:defK} 
\end{align} 
where we have introduced Majorana modes $\chi_{m,n}$ as $\phi_{c,n} = (\chi_{2c-1,n} + \mathrm{i} \chi_{2c,n})/\sqrt{2}$ to make the $\mathsf{O}(2N)$ symmetry of this lattice operator explicit.

Before adding the GN four point interaction, let us first discuss how to add an explicit mass term. Having fixed $\alpha = \gamma^0\gamma^1 = \sigma^x$, the field theory allows for any choice $\beta =\gamma^0= \cos(\theta) \sigma^y + \sin(\theta) \sigma^z$. The original proposal of Susskind in Ref.~\cite{susskind1977lattice} was $\beta=\gamma^0 = \sigma_z$, which is then trivially discretised into a lattice mass term on the doubled lattice as
\begin{equation}
\Delta \sum_n (-1)^n \sum_{c = 1}^{N} \phi_{c,n}^\dagger \phi_{c,n}\label{eq:susskindmass}	
\end{equation}
with $\Delta = m a$ the mass in dimensionless lattice units.
This clearly indicates how one-site translations on the staggered lattice flip the sign of the mass term, and can thus be related to a lattice remnant of the discrete chiral transformation $\psi \to \gamma^5\psi$. However, this lattice term breaks the $\mathsf{O}(2N)$ symmetry, as can be made explicit by rewriting it in terms of the Majorana components. 

The alternative choice $\beta = \sigma^y$ yields terms involving both components of the Dirac spinor on the same position, which can be discretised on our staggered lattice by averaging one of the two components over the two nearby positions. This gives rise to an alternative lattice mass term, which takes the form of a staggered hopping
\begin{align}
\frac{\Delta}{2} \sum_n (-1)^{n} K_{n,n+1}\label{eq:sshmass}	
\end{align}
and thus respects the $\mathsf{O}(2N)$ symmetry. This term is well known from the SSH model, where the alternating hopping strengths result from dimerisation. 

On the lattice, these two mass terms, resulting from two different choices for $\beta$ (and thus, ultimately, a different choice of basis for the Dirac spinor in the continuum) are not equivalent. From the periodic table of topological insulators and superconductors\cite{schnyder2008classification,kitaev2009periodic}, it is well known that the SSH mass term preserves sublattice symmetry (class AIII or BDI), which gives rise to a protected topological invariant labeled by $\mathbb{Z}$. Sublattice symmetry is also known as chiral symmetry in that context, but we refrain from using this terminology, as it is clearly different from the chiral symmetry of the field theory relevant to our study, and which is broken by either mass term.

Writing the Hamiltonian terms in momentum space after blocking two sites, they take the form 
\begin{align}
	\sum_{c=1}^{N} \int_{-\pi}^\pi  \Psi_c(p)^\dagger  \del{\vec{d}(p) \cdot \vec{\sigma}} \Psi_c(p)\,\mathrm{d}p 
\end{align}
similar to the field theory Hamiltonian but with $\vec{d}(p)$ a periodic function of the lattice momentum $p \in [-\pi,+\pi)$ on the blocked lattice. A gapped model has nonzero $\vec{d}(p)$ for all $p$. Sublattice symmetry imposes that $\vec{d}(k)$ is confined to a two-dimensional space, and the topological invariant corresponds to the winding number of $\vec{d}(p)$ around the origin. Both the kinetic term in Eq.~\eqref{eq:dirackinetic} and the SSH mass term in Eq.~\eqref{eq:sshmass} only have non-zero $d_x$ and $d_y$ components, in particular $d_x(p) = \sin(p) (1 + \frac{\Delta}{2})$ and $d_y(p) = (1+\frac{\Delta}{2}) + \cos(p) (1 - \frac{\Delta}{2})$. They lead to a well-defined winding number, which is non-zero for $\Delta <0$ (shifting the unit cell definition is equivalent to $\Delta \rightarrow -\Delta)$, thus indicating a symmetry-protected topological (SPT) phase protected by either sublattice symmetry or bond-centred inversion, where the latter is also defined for interacting systems. Susskind's mass term [Eq.~\eqref{eq:susskindmass}] corresponds to $d_z(p) = \Delta$ and breaks the topological invariant. In the field theory, the kinetic term has a single component (i.e.\ $d_x(p) = p $ if $\alpha = \sigma^x$), and so either choice of $\beta$ is equivalent. While the winding number is undefined as momentum space is unbounded, topological features still manifest themselves when considering a domain between positive and negative mass, which gives rise to gapless edge modes, as described by Jackiw and Rebbi \cite{jackiw1976solitons}. Hence, the SSH mass term provides a more faithful lattice description of the massive Dirac field.

\begin{figure}[t!]
	\centering
	\begin{tikzpicture}
		\draw[black, thick       ] (-4,0) -- (    0,0);
		\draw[black, thick, ->] (-4,0) -- (-0.1,0);
		\draw[black, thick, ->] (-4,0) -- (-0.2,0);
		
		\draw[black, thick,       ] (0,0  ) -- (4,0);
		\draw[black, thick, <- ] (0.1,0.) -- (4,0);
		\draw[black, thick, <- ] (0.2,0.) -- (4,0);

		\draw[red, thick,      ] (0,0) -- (0,2);
		\draw[red, thick, <-] (0,0.1) -- (0,2);
		\draw[red, thick, <-] (0,0.2) -- (0,2);
		
		\node at (-2,1) {$\mathsf{O}(2N)$ SPT phase};
		\node at (2,1) {trivial phase};
		\node at (0,-0.35) {$c=N$};
		
		\filldraw[black] (0,0) circle (2pt) ;
		
		\node at (4,0.2) {$\Delta$};
		\node at (0.2,2) {$g$};
	\end{tikzpicture}
	\caption{Phase diagram of the lattice GN Hamiltonian from Eq.~\eqref{eq:gnlattice} with additional SSH mass term from Eq.~\eqref{eq:sshmass}. The point $g=\Delta=0$ is the lattice realisation of the $N$-flavour free fermion conformal field theory, which plays the role of a UV fixed point. For negative/positive mass perturbations the ground state is in a symmetry protected/trivial gapped phase respectively. The GN model has no explicit mass term and corresponds to a first order phase transition between those phases, i.e.\ it has two gapped ground states with zero and non-zero topological invariant respectively. The continuum limit of the Gross Neveu model approaches the $c=N$ point via the red arrows, while the black arrows denote the continuum limit of $N$ massive fermions.}
	\label{fig:phasediagram}
\end{figure}
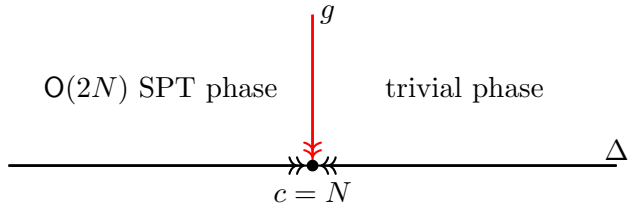

We can rewrite the mass term from Eq.~\eqref{eq:sshmass} as
\begin{equation}
	\frac{\Delta}{2} \sum_{n} (-1)^n K_{n,n+1} = \frac{\Delta}{2} \sum_{n} (-1)^{n} \Sigma_{n,n+1,n+2}\label{eq:sshmassalt}
\end{equation}
with the three-site operator
\begin{equation}
\Sigma_{n,n+1,n+2} =  \frac{K_{2n,2n+1}-K_{2n+1,2n+2} }{2},\label{eq:defSigma}
\end{equation}
which plays the role of a local order parameter, i.e.\ the lattice equivalent of $\bar{\psi}\psi$. Whereas $K_{n,n+1}$ has non-zero expectation value even with respect to the $\Delta=0$ ground state, $\Sigma_{n,n+1,n+2}$ is an absolute measure for the mass condensate. With this, it has now become straightforward to formulate a lattice Hamiltonian for the GN model,
\begin{align}
aH = \sum_n \del{ K_{n,n+1} - \frac{g^2}{4}\Sigma_{n,n+1,n+2}^2 },
\label{eq:gnlattice}
\end{align}
where the interaction coefficient was changed from $g^2/2$ to $g^2/4$ as we associated one interaction term with every site of the doubled lattice. Doing so, this model has single-site translation invariance, which  corresponds to the lattice remnant of discrete chiral symmetry, as well as $\mathsf{O}(2N)$ symmetry. The projective nature of the $\mathsf{O}(2N)$ action on the single-site Hilbert space, which is discussed in Sec.~\ref{sec:mps}, in combination with translation invariance enables the application of the Lieb-Schultz-Mattis theorem \cite{LiebSchultzMattis}: this model cannot have a unique, gapped ground state. It is critical for $g=0$, but we expect the interaction to be marginally relevant and lead to a symmetry broken state for $g\neq 0$. The Mermin-Wagner theorem excludes the $\mathsf{O}(2N)$ symmetry to be broken, thus leading to dimerisation, the lattice manifestation of a mass condensate, as the most likely scenario. By adding an explicit SSH mass term, a two-dimensional phase diagram is obtained, depicted in Fig.~\ref{fig:phasediagram}, where the Hamiltonian in Eq.~\eqref{eq:gnlattice} (\textit{i.e.}~the $\Delta=0$ line) can be identified as a first order phase transition between the trivial and non-trivial SPT phase. While this is somewhat similar to the $\mathbb{Z}_2$-symmetry breaking phase of the Ising model being a first order line between the the explicit symmetry-broken regimes with positive and negative longitudinal field, the topologically distinct nature of the phases at both sides of the transition results in symmetry fractionalisation in the kink excitations that interpolate between the two ground states at the first order line. 

We confirm that the low-energy behaviour of this lattice model does indeed replicate all features of the Gross-Neveu field theory in the next sections using a large $N$ calculation (Section~\ref{sec:meanfield}) and by constructing an MPS ansatz (Sections~\ref{sec:mps} and \ref{sec:MPSgap}). The MPS ansatz is not only  used for numerical simulations for $N=2$, but also provides further insight into the symmetry structure of the excitation spectrum for general $N$.

\section{Large-N solution}
\label{sec:meanfield}

Similar to the original paper of Gross and Neveu \cite{gross1974dynamical}, we first study our lattice model in the limit of $N\to \infty$, but within the Hamiltonian formalism. In the limit $N\to\infty$, the permutation symmetry corresponding to exchanging the different flavours in combination with monogamy of entanglement \cite{fannes1988symmetric,terhal2004entanglement} can be used to argue that the ground state $\ket{\Psi}$ will be a product state over the different flavours, where each flavour is described by the same state: $\ket{\Psi} = \ket{\phi}^{\otimes N}$. The energy of this state is given by
\begin{align}
	E_\Psi =&  \sum_{n} \del{ N\braket{\phi| k_{n,n+1} - \frac{g^2}{4} \sigma_{n,n+1,n+2}^2 |\phi} - \frac{g^2}{4} N (N-1)  \braket{\phi|\sigma_{n,n+1,n+2}|\phi}^2 }
\end{align}
with $k_{n,n+1}$ and $\sigma_{n,n+1,n+2}$ the single-flavour versions of $K_{n,n+1}$ and $\Sigma_{n,n+1,n+2}$ respectively. The GN interaction splits into $N$ terms which act on a single flavour, and $N(N-1)$ terms which act across different flavours, and are transformed into a product of expectation values due to our product state ansatz: correlations between flavours vanish for $N\to \infty$. In order to obtain finite results, we take the limit $N\to \infty$ while keeping $\lambda = g^2(N-1)$ fixed. As $g^2$ itself goes to zero, the self-interaction of the flavours vanishes.

Minimising the energy with respect to $\phi$ yields, after adding a Langrange constraint for the normalisation, a self-consistent eigenvalue problem for the state $\ket{\phi}$:
\begin{align}
	\sum_n \del{ k_{n,n+1} - \frac{\lambda\braket{\sigma_{n,n+1,n+2}}}{2}\sigma_{n, n+1, n+2}  } \ket{\phi} = E_\phi \ket{\phi}
\end{align}
 where the Lagrange parameter $E_\phi$ can be interpreted as the energy of a single flavor, and 
 \begin{align}
 	E_\Psi = N (E_\phi + \frac{\lambda}{4} \sum_{n} \braket{\phi|\sigma_{n,n+1,n+2}|\phi}^2)
 \end{align}
 
Assuming a dimerised solution, we set $\braket{\phi|\sigma_{n,n+1,n+2}|\phi} = (-1)^{n} \sigma_0$. The state $\ket{\phi}$ of a single flavor is now determined as the ground state of the quadratic mean-field (\textit{i.e.}\ Hartree-Fock) Hamiltonian
\begin{equation}
	H_{\text{MF}} = \sum_{n} k_{n,n+1} + \frac{\lambda \sigma_0}{2} (-1)^{n} \sigma_{n,n+1,n+2}
\end{equation}
in which $\lambda \sigma_0$ plays the role of an SSH mass. The mean-field Hamiltonian is diagonalised by blocking the lattice and going to momentum space, which gives rise to single particle energies
\begin{align}
	\varepsilon(p) = \pm\sqrt{4\sin^2(p/2) + \lambda^2\sigma_0^2\cos^2(p/2) }.
\end{align}
with, as before, $p\in[-\pi,+\pi)$ the lattice momentum on the blocked lattice.
One can verify that the self-consistency condition for $\sigma_0$ is equivalent to minimising $E_\Psi/N = \braket{\phi|H_{\text{MF}}|\phi} + \sum_{n} \frac{\lambda}{4} \sigma_0^2$, or thus
\begin{align}
	\frac{e_\Psi}{N} = &\frac{\lambda}{2}\sigma_0^2 -\int_{-\pi}^{\pi} \frac{\mathrm{d}p}{2\pi} \sqrt{4\sin^2(p/2) + \lambda^2\sigma_0^2\cos^2(p/2) }
\end{align}
with $e_\Psi$ the energy density associated with the sites of the blocked lattice. The value of $\sigma_0$ is thus determined by the condition
\begin{align}
	\frac{1}{\lambda} = \int_{-\pi}^{\pi}\frac{\mathrm{d}p}{2\pi}\frac{\cos^2(p/2)}{\sqrt{4\sin^2(p/2)+\lambda^2\sigma_0^2\cos^2(p/2) }}.
\end{align}
This can be further expanded as
\begin{align}
	\frac{1}{\lambda} = \frac{1}{\pi} \frac{K\del{1- \frac{\lambda^2\sigma_0^2}{4}} - E\del{1 - \frac{\lambda^2\sigma_0^2}{4}}}{1 - \frac{\lambda^2\sigma_0^2}{4}}
\end{align}
with $K$ and $E$ the complete elliptic integral of the first and second kind, respectively. An asymptotic expansion for small $\lambda \sigma_0$, which is the dimensionless mass and should go to zero to recover the continuum limit, yields
\begin{align}
	\frac{1}{\lambda} =& -\frac{1}{2\pi} \left[\log\del{\frac{\lambda^2 \sigma_0^2}{64}} + 2\right] + O\del{\lambda^2\sigma_0^2 \log\del{\lambda^2\sigma_0^2}  }
\end{align}

As a result, the effective fermion mass is in the large-$N$ limit given by
\begin{equation}
	a m_1 = \lambda \sigma_0 = \frac{8}{\mathrm{e}} \exp\left[-\frac{\pi}{(N-1)g^2}\right] \del{1+\mathcal{O}(g^2)}\label{largeN-mass}
\end{equation}
such that $m_1$ is indeed proportional to the RG-invariant mass scale $\Lambda$ that was introduced in Section~\ref{sec:GN_nut}, with $\mu = a^{-1}$. In particular, by comparing to the $N\to \infty$ limit of the exact result in Eq.~\eqref{eq:exactmassgap}, i.e.\ $m_1 = \Lambda_{\overline{\text{MS}}}$, we are lead to conclude that if we define
\begin{equation}
	\Lambda_{\text{lat}} = \frac{1}{a} \left(\frac{(N-1)g^2}{2\pi} \right)^{\frac{1}{2N-2}} \exp\left[-\frac{\pi}{(N-1)g^2}\right]
\end{equation}
then $\Lambda_{\text{lat}} = \frac{8}{e} \Lambda_{\overline{\text{MS}}}$ and thus we should recover
\begin{equation}
	m_1/\Lambda_{\text{lat}} = \frac{8}{e} \frac{(4e)^{\frac{1}{2N-2}}}{\Gamma(1 - \frac{1}{2N-2})}(1 + O(g^2))
\end{equation}
in the continuum limit $g\to 0$. However, the $N\to \infty$ solution is in itself not sufficient to support this conclusion, as other $N$ dependent scale factors might appear. A careful comparison between $\Lambda_{\text{lat}}$ and $\Lambda_{\overline{\text{MS}}}$ using the fermion-fermion scattering amplitude at finite $N$ leads to the same result, as explained in Appendix~\ref{sec:appendix}.

Henceforth, we omit the lattice spacing $a$, as it appears trivially in length or mass scale quantities and does not directly affect the distance to the continuum limit. So we stop differentiating between dimensionless lattice and field theory quantities.

\section{Matrix product states}
\label{sec:mps}
For finite $N$, correlations between the different flavours cannot be ignored, and the ground state of our lattice model is a fully correlated quantum state, both in the spatial and in the flavour direction. We now try to approximate this ground state using a MPS ansatz, which is known to capture the quantum correlations in low-energy states of gapped local Hamiltonians for quantum spin chains \cite{hastings}. The MPS ansatz associates with every site $n$ of such a spin chain a 3-leg tensor $A(n)$ of size ${D_{n-1}\times d \times D_{n}}$, with $d$ the local Hilbert space dimension of the physical index. The left (respectively right) virtual index of size $D_{n-1}$ ($D_{n}$) is then contracted with the right virtual index of the previous (left virtual index of the next) tensor, resulting in a correlated state whose bipartite entanglement for a cut between site $n$ and site $n+1$ is upper bounded by $\log(D_n)$, independent of the system size (in accordance with the area law for entanglement entropy in one-dimensional systems \cite{arealaw}). By defining a unit cell, i.e.\ a periodic $n$ dependence in the tensors $A(n)$, we can describe quantum states directly in the thermodynamic limit.

Our lattice model is easily translated into a spin chain using a Jordan-Wigner transformation
\begin{equation}
	\phi_{c,n} = \del{\prod_{n'<n}\prod_{c'} \sigma^{z}_{c',n'}} \del{\prod_{c'<c} \sigma^{z}_{c',n}} \sigma^{-}_{c,n}
\end{equation}
where we introduce a linear ordering in the flavour direction $c=1,\ldots,N$, and thus associate $N$ qubits or spins with each site, so that the local Hilbert space dimension is $2^N$. We keep these $N$ qubits together (as opposed to treating them as $N$ individual sites) in order to preserve the $\mathsf{O}(2N)$ symmetry and to be able to capture it in the MPS ansatz. Using this particular ordering in the Jordan-Wigner transformation, the generators of the associated Lie algebra $\mathfrak{so}(2N)$ transform into a sum of one-site operators so that the resulting symmetry transformations act on-site. The local Hilbert space of a site can be identified with the direct sum of the two fundamental $\mathfrak{so}(2N)$ spinor representations (each of which has dimension $2^{N-1}$) with opposite total fermion parity.

In order to construct $\mathsf{SO}(2N)$ symmetric MPS\footnote{By working with the representations of the Lie-algebra $\mathfrak{so}(2N)$, we effectively only impose $\mathsf{SO}(2N)$ symmetry. However, we discuss the role of the additional mirror symmetry that extends $\mathsf{SO}(2N)$ to $\mathsf{O}(2N)$ for the particular case $N=2$ which was used in our simulations, and find that it is unbroken.}, the virtual indices of the tensors should also carry representations of the group \cite{symmMPS1, symmMPS2} and the local tensors $A(n)$ should intertwine the representation on the right virtual index with the tensor product of the representations on left virtual and physical index. The representation on the physical index, i.e.\ the direct sum of the two fundamental spinor representations, is projective\footnote{The spinor representations of $\mathsf{(S)O(N)}$ are linear representations of the universal covering group, known as $\mathsf{(S)Pin}(N)$.}. Therefore, if the right virtual index is associated with a linear representation (i.e.\ a direct sum of tensor representations of $\mathsf{SO}(2N)$), then the left virtual index should also be projective (and thus be composed of spinor irreducible representations), and vice versa. We are thus naturally led to a two-site unit cell. This is the MPS manifestation of the Lieb-Schultz-Mattis theorem \cite{sanz2009mps}: MPS represent finitely correlated (and thus gapped) states, and cannot be simultaneously invariant under translation symmetry and an on-site symmetry whose physical action is projective. As the on-site symmetry is continuous and cannot be broken, we thus propose an ansatz for the ground state with a two-site unit cell, in line with the expected dimerisation:
\begin{align}
\ket{\psi[A_1, A_2]} &= \sum_{\vec{s}} \del{\prod_{n\in\mathcal{Z}} A_{1,s_{2n}} A_{2,s_{2n+1}} \ket{s_{2n}s_{2n+1}}}   \label{MPS} = \diagramw{5cm}{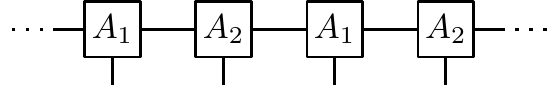} 
\end{align}

We can also formulate MPS-based ans\"atze for elementary excitations on top of the ground state \cite{haegeman2012variational,zauner2018topological, medley}, as well as kink excitations that interpolate between the two ground states. Both topologically trivial excitations and kinks can be created by modifying a single tensor (which has an effect on an extended region) and building a proper momentum superposition (unlike in semiclassical studies where the kinks or solitons are localised in real-space). The ansatz for kinks, where the two different ground states (corresponding to a one-site shift of the unit cell) surround the new tensor, is diagrammatically represented as
\begin{align}
\ket{K_p} &= \sum_n\del{e^{ipn}\diagramwt{5cm}{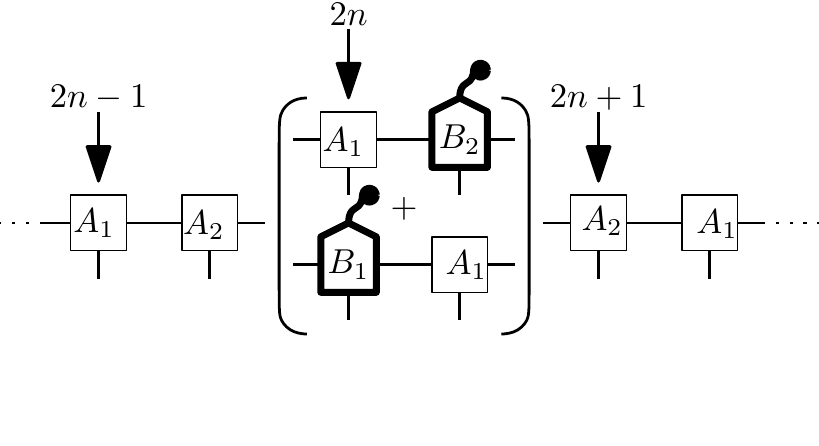}}
\end{align}
with once again $p\in [-\pi,+\pi)$ the momentum on the blocked lattice. The new tensors (labelled $B_i$) carry an additional leg corresponding to the Hilbert space of the irreducible representation of the excitation that is being targeted. In the case of kinks, as depicted here, the two virtual legs of $B$ tensors need to be identical and thus carry the same representations, and it follows automatically that physical symmetry sector of the kink states needs to be a spinor representation. These constructions are well-known in the case of half-integer spin chains \cite{laurens2020}, where they also correspond to the renowned result that the elementary excitations in e.g.~the half-integer spin Heisenberg chain are spinors \cite{faddeev1981spin, shastry1981excitation}. The analogous construction for topologically trivial excitations on top of a single ground state illustrates that these are labelled by linear irreducible representations of $\mathsf{SO}(2N)$. For an in-depth review on these excited states and their implementation we refer to Ref. ~\cite{medley}.

The variational ansatz for excitations gives rise to an energy-momentum dispersion relation, e.g. $E_K(p)$ for the kink state $\ket{K_p}$, from which we can extract a range of mass scales related to its value, inverse curvature and higher derivatives at $p=0$, where the dispersion relation has its minimum. Indeed, by fitting 
\begin{align}
E_{K}(p) = m_{K,1}\del{1+\frac{p^2}{2m_{K,2}^2}+\cdots}
\end{align}
to the dispersion relation for small values of the lattice momentum $p$, we obtain two different mass scales\footnote{One could expand further and introduce an arbitrary amount of mass scales.}. Alternatively, expecting relativistic invariance, we can rewrite this expansion for the square of the energy as
\begin{align}
E_K(p)^2 = m_{K,1}^2 + \frac{m_{K,1}^2}{m_{K,2}^2} p^2 + \cdots
\end{align}
and thus interpret the ratio $m_{K,1}/m_{K,2}$ as an effective speed of light, which should go to one if Lorentz invariance is obtained in the continuum limit.

In a theory near a relativistic continuum limit, information about the particle masses is also encoded in the ground state, more particularly in the spectrum of inverse correlation lengths. This is true for trivial excitations, by writing two-point correlation functions using the K\"{a}ll\'{e}n-Lehmann representation \cite{zauner2015transfer}. To extract the kink mass from a correlation function, one needs to study the correlation of string operators. In the MPS language, the corresponding inverse correlation length can easily be extracted by studying the \emph{mixed} transfer matrix, made from two different ground states. In this particular case, these two ground states are related by a one-site shift, and we define
\begin{align}
m_{K,3} &= -\log\del{\rho \del{\diagramw{2cm}{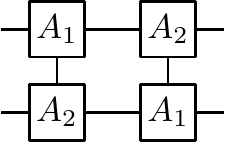}}} \;,
\label{corlen}
\end{align}
with $\rho$ the spectral radius (largest magnitude eigenvalue). The right hand side of this equation gives an inverse length scale, and requires a (dimensionless) velocity to give an energy scale. Again we assume this velocity to be one as we approach the continuum limit, and we will directly compare $m_{K,3}$ as another estimate of the kink mass. Note that the eigenvectors of the mixed transfer matrix, due to the different nature of the two legs on which they act, also transform according to spinor representations, reflecting the spinorial nature of the kinks to which the mixed transfer matrix is related. One could also calculate the leading eigenvalue in the trivial transfer matrix as an estimate for the fermionic correlation length an hence inverse fermion mass.

While (inverse) correlation lengths converge slowly as a function of the MPS bond dimension, a scaling theory for their behaviour was recently developed based on a parameter $\delta$ that quantifies the level spacing in the logarithmic eigenvalue spectrum of the transfer matrix (which should become continuous in the infinite bond dimension limit). We refer to Refs.~\cite{rams2018precise,vanhecke2019scaling} and use these techniques to extrapolate reported mass values to the infinite bond dimension limit.

\section{Simulation of the $N=2$ mass gap}
\label{sec:MPSgap}
As a first application of our MPS simulations, we compute the different estimations of the mass gap $m_K$ of the kink that interpolates between the two vacua for $N=2$. Hereto, we first find the optimal MPS representation of the ground state using the ``variational algorithm for uniform matrix product states'' (VUMPS)\cite{vumps}, which directly minimises the energy (density) (\textit{i.e.} variationally) in the thermodynamic limit.

Our implementation of this algorithm, as well as the algorithm for computing the dispersion relation using the excitation ansatz, can be found in ``MPSKit.jl"\cite{mpskit}, an open source package for MPS algorithms using the scientific programming language Julia. This package builds upon ``TensorKit.jl" \cite{tensorkit}, a lower level open source package for representing and manipulating tensors with arbitrary (abelian and non-abelian) symmetries.

Specifically for $N=2$,  we enforced the tensors to be representations of $\mathsf{SO(4)}$, or rather its universal cover $\mathsf{Spin}(4)$. This group  is equivalent to $\mathsf{SU}(2)\times \mathsf{SU}(2)$ and irreducible representations are labeled by a tuple of two $\mathsf{SU}(2)$ quantum numbers, \textit{i.e.}\ half integers or integers. The resulting representation is a projective (\textit{i.e.}\ spinor) representation of $\mathsf{SO}(4)$ if only one of both quantum numbers is a half-integer. With both quantum numbers integer or half-integer, a linear (i.e.\ tensor) representation of $\mathsf{SO}(4)$ is obtained. This symmetry can easily be understood by considering the two sets of generators,
\begin{align}
	S^+ &= (S^-)^\dagger= \sum_{n}\phi_{1,n}^\dagger \phi_{2,n}\\
	S^z &= \sum_{n} \frac{\phi_{1,n}^\dagger\phi_{1,n} - \phi_{2,n}^\dagger\phi_{2,n}}{2}\\
	T^+ &= (T^-)^\dagger = \mathrm{i}\sum_{n} \phi_{1,n}^\dagger \phi_{2,n}^\dagger\\
	T^z &=\sum_{n} \frac{\phi_{1,n}^\dagger\phi_{1,n} + \phi_{2,n}^\dagger\phi_{2,n}-1}{2}
\end{align}
corresponding to rotations in flavour space (odd fermion subspace of single occupancy) and in some pseudospin space (even fermion subspace of zero or double occupancy), exactly as in the Hubbard model at half filling \cite{Essler}. The disconnected part of $\mathsf{O}(4)$ (or its double cover, $\mathsf{Pin}(4)$) is generated by a fermionic particle-hole transformation on one of the fermion flavours, and has the effect of interchanging the two $\mathsf{SU}(2)$ factors. It thus results in degeneracies between sectors $(j_1,j_2)$ and $(j_2,j_1)$, i.e.\ whenever $j_1\neq j_2$, these two representations will always come together as $(j_1,j_2) \oplus (j_2,j_1)$, where the direct sum constitutes a proper (linear or projective) representation of $\mathsf{O}(4)$. While this extra symmetry is not enforced on our MPS representation, it does seem to be perfectly preserved in the ground states we find numerically. Exploiting the $\mathsf{Spin}(4) \cong \mathsf{SU}(2)\times \mathsf{SU}(2)$ symmetry has enabled us to push the bond dimension to $D \approx 4000$. 

Using the exact equation for the mass gap [Eq.~\eqref{eq:exactmassgap}], the relation between $\Lambda_{\overline{\text{MS}}}$ and $\Lambda_{\text{lat}}$ proven in Appendix~A, and the fact that the kink mass is half the fermion mass for $N=2$, we obtain that the dimensionless mass of the elementary kinks for $N=2$ should approach the value
\begin{align}
	m_K = \frac{8}{e}\sqrt{\frac{e}{\pi}}\frac{1}{\sqrt{2\pi}}g e^{-\pi/g^2} \label{eq:N2mass}
\end{align}
in the continuum limit.

\begin{figure}[t!]
	\centering
	\includegraphics[width=0.49\columnwidth]{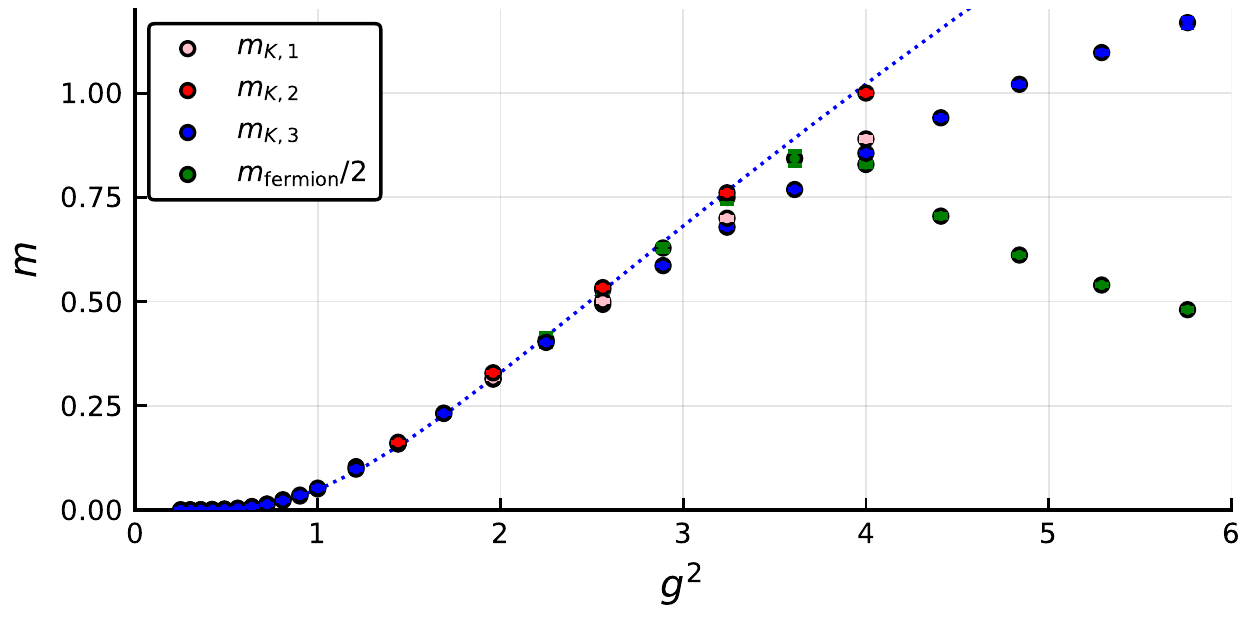}
	\includegraphics[width=0.49\columnwidth]{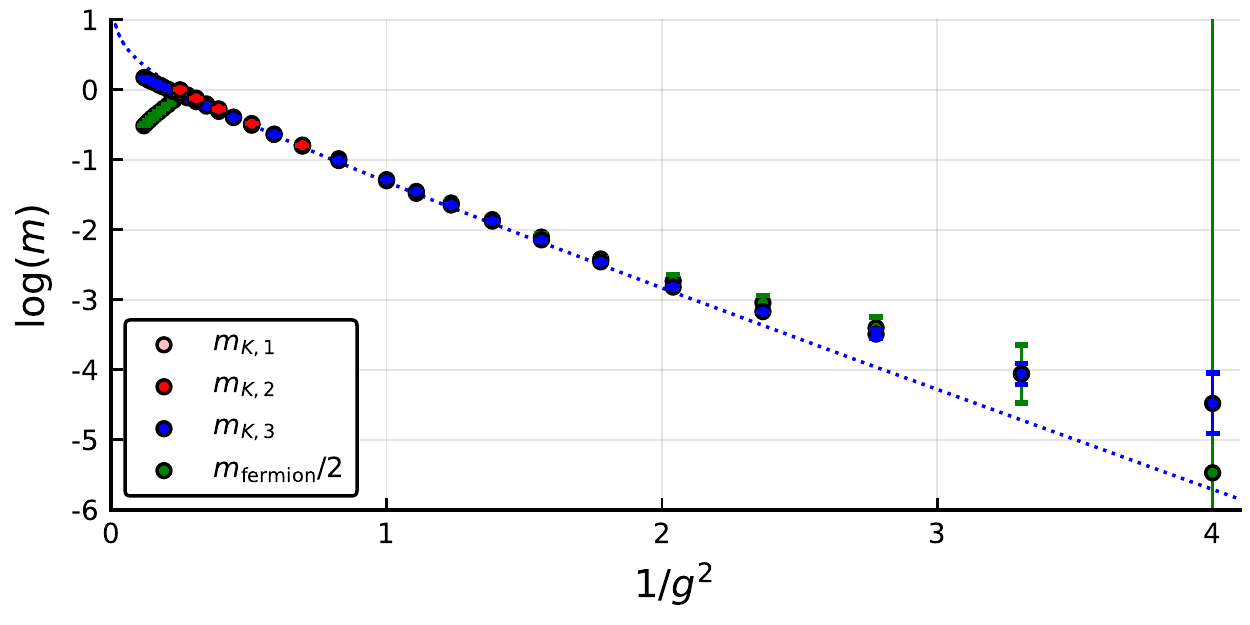}
	\caption{Extrapolated mass scales and a fit to small couplings, presented in two different ways. The linear behaviour in the second panel follows from the leading order contribution in $\log(m_K)$. The error bars correspond to the uncertainty in the $\delta \rightarrow 0$ extrapolations. Fits are made against the inverse correlation length ($m_{K,3}$) data and take the form $\log(m_{K,3}) = \log(C \Lambda_{\text{lat}})$ and $C$ is consistent with the value predicted by the field theory results up to $4\%$.  }
	\label{relativity_check}
\end{figure}

\begin{figure}[t!]
	\centering
	\includegraphics[width=0.6\columnwidth]{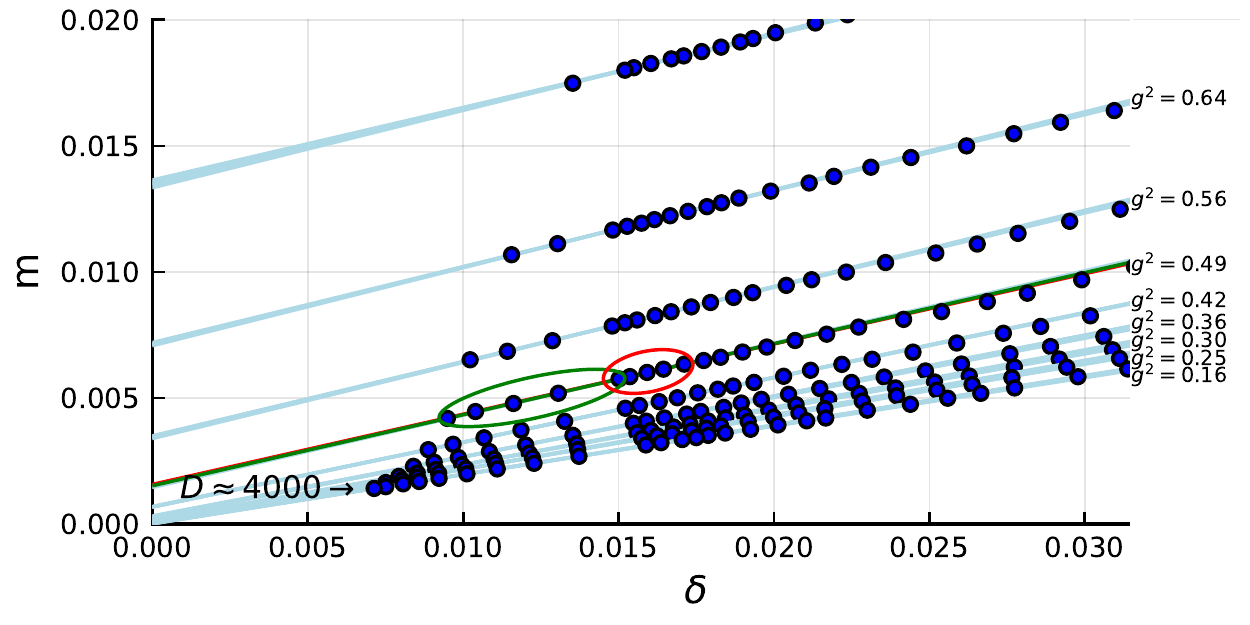}
	\caption{Extrapolation of the inverse correlation length of the mixed transfer matrix (topological sector), corresponding to the mass estimator $m_{K,3}$, as a function of $\delta$, the spacing in the logarithmic spectrum of the transfer matrix \cite{rams2018precise}. For every value of the coupling $g^2$ we show a handful of linear extrapolations towards $\delta=0$ each taking different points into consideration. For example at $g^2=0.56$ we have highlighted the 5 points with highest bond dimension and the corresponding extrapolation in green. Another extrapolation where we discarded the 4 points with highest bond dimension is highlighted in red. These extrapolate to slightly different masses which allows us to estimate the error on the extrapolation that is shown in  Fig.~\ref{relativity_check}.    }
	\label{extrapol}
\end{figure}

Fig.~\ref{relativity_check} depicts four extrapolated mass scales as a function of the coupling. The blue dots, referred to as $m_{K,3}$, \textit{i.e.}~the inverse correlation length extracted from the mixed transfer matrix, are relatively cheap to compute but require careful extrapolation towards $\delta = 0$. This extrapolation is illustrated in Fig.~\ref{extrapol}. Here,  we present a handful of linear extrapolations for each coupling. One fit considers only the 5 highest bond dimension simulations, other extrapolations discard the highest bond dimensions and use the 5 next best bond dimensions. The resulting variation in the extrapolated mass gap gives rise the error bars depicted in Fig.~\ref{relativity_check}. In a similar manner, the largest correlation length extracted from the normal transfer matrix (topologically trivial sector) should be determined by the fermion mass, and half of this value should also provide an estimate of the kink mass in the continuum limit. It is depicted by the green dots in Fig.~\ref{relativity_check}. Note that for large values $g \gtrsim 4$ (which are not relevant for the continuum limit), the fermion mass (as extracted from the inverse correlation length), is less than twice the kink mass. Hence, at those values of the coupling constant, the lattice model is likely to exhibit a stable particle in the topologically trivial sector.

\begin{figure}
	\centering
	\includegraphics[width=0.6\columnwidth]{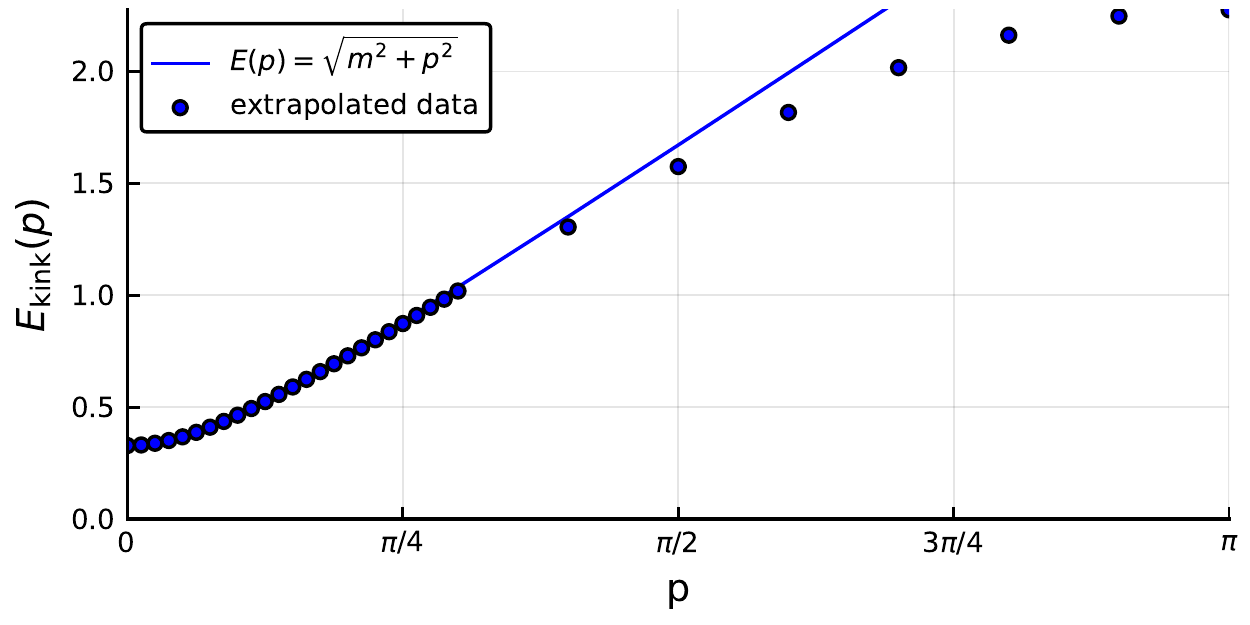}
	\caption{The kink energy as a function of the momentum on the blocked lattice for $g=1.4$, as compared to the corresponding Einstein energy-momentum relation, indicating Lorentz invariance over a relatively large range of momenta.}
	\label{dispersion}
\end{figure}

The pink and red dots in Fig.~\ref{relativity_check} correspond to the gap and inverse curvature obtained from the excitation ansatz, referred to as $m_{K,1}$ and $m_{K,2}$ before. These points converge more quickly with bond dimensions and require little extrapolation, yet are more costly to obtain. We have only calculated these points for a selection of couplings. The observation that the value of the gap, its curvature and the inverse correlation length coincide clearly shows an emergent Lorentz symmetry with speed of light equal to one, as intended. To further illustrate this, we plot the kink dispersion relation at $g=1.4$ in Fig.~\ref{dispersion} and compare the dispersion to the relativistic prediction $E_p^2 = m^2 + p^2$. Even for relatively large lattice momenta up to $p \approx \pi/2$ the correspondence is good. Note that the mass here is already of the order of $0.4$ in lattice units, and we are thus already quite far from the proper continuum limit. For even larger values of $g^2$, where the mass becomes of the order of one in lattice units, deviations between the different mass scales can be observed, as expected.

To compare our mass data to the proposed continuum limit of Eq.~\eqref{eq:N2mass}, it is useful to consider the right panel of Fig.~\ref{relativity_check} where $\log(m_K)$ is displayed as a function of $1/g^2$. For sufficiently small couplings the logarithm is dominated by $-\pi/g^2$ resulting in a linear relation that is ideal for fitting.  A fit of the form $\log(m_K)  = -\frac{\pi}{g^2} + \log(g) + \log\del{C}$ is shown in both panels and we find $C \approx 1.121$ for the inverse correlation length data $m_{K,3}$. This value compares well with the expected result $C=\frac{8}{e}\sqrt{\frac{e}{\pi}}\frac{1}{\sqrt{2\pi}} = 1.092$, but shows a small overshoot of about $3\%$. Note that the logarithmic contribution to the fit complicates the fitting process and makes the fit parameter very sensitive to data in the small $g$ regime, where the masses become extremely small and thus hard to pinpoint exactly. Nonetheless, we do conclude that this data provides ample evidence for the continuum limit of our lattice model being well described by the GN field theory, as intended.

\section{Entanglement structure of the groundstate}
\label{sec:entanglement}

Another advantage of tensor network representations of quantum states is that they give full access to the entanglement structure of the state, which is an interesting concept in its own right as it provides an fresh perspective into the quantum correlations of the state and has thus received a lot of attention lately. In particular, the half-space reduced density matrix defines the entanglement Hamiltonian $H_{E}$ via
\begin{align}
\text{Tr}_{\text{half space}}\del{\ket{\Psi}\bra{\Psi}} =  \hat\rho = e^{-2\pi H_{E}} \label{red_density}
\end{align}
which for a relativistic field theory is also known as the modular or Rindler Hamiltonian (corresponding to an accelerating observer). For a conformal field theory (CFT), the modular Hamiltonian can be mapped back to the original Hamiltonian using a conformal (logarithmic) transformation. In a gapped theory close to a CFT in the ultraviolet (UV), we expect the entanglement structure, which is anyway determined by UV modes of the theory \footnote{This is an example of a ultraviolet-inrared duality; the lowest entanglement modes (dominant singular values) are related to short-range degrees of the freedom, the infinite abundance of which causes the divergence of the entanglement entropy in the continuum limit.}, to follow the CFT prediction closely up to length scales of the correlation length. The logarithmic mapping should thus transform the modular Hamiltonian onto the CFT Hamiltonian on a finite system with length approximately given by the logarithm of the correlation length in the system. This argument was recently formalised for CFTs perturbed by a relevant interaction by Cho, Ludwig and Ryu \cite{universal-entanglement}. In what follows we we will first calculate the prediction for the entanglement spectrum for general $N$. We will then check that the prediction matches our simulations for $N=2$ despite the marginal nature of our interaction term.

\subsection{General result}

Anticipating that the entanglement spectrum corresponds to the CFT spectrum on a finite system with open boundary conditions, we thus compute the spectrum for $N$ free fermions (the UV fixed point of our model) on a finite interval of length $L$. As before, we denote with $\lambda_m$ ($m=1,\ldots,2N$) the $2N$ Majorana fields, and with $\lambda_{m,1}$ and $\lambda_{m,2}$ their two spinor components. After partial integration, the CFT Hamiltonian is given by
\begin{equation}
	H_{E} = \int_0^L \sum_{m\in2N} 2i \lambda_{m,1}\partial_x \lambda_{m,2}
\end{equation}
Conformal boundary conditions can be of the type
\begin{equation}
\lambda_{m,1}|_0 = \lambda_{m,2}|_L = 0 \hspace{1cm}\text{and}\hspace{1cm} \partial_x\lambda_{m,2}|_0 = \partial_x\lambda_{m,1}|_L = 0
\end{equation}
which results in a standing wave expansion with half-integer momenta
\begin{align}
	\begin{cases}
		\lambda_{m,1}(x) = \sqrt{\frac{2}{L}}\sum_{k>0} \hat{\lambda}_{m,1}(k)\sin\del{\frac{\pi\del{k-1/2}x}{L}}\\
		\lambda_{m,2}(x) = \sqrt{\frac{2}{L}}\sum_{k>0} \hat{\lambda}_{m,2}(k)\cos\del{\frac{\pi\del{k-1/2}x}{L}}
	\end{cases},
\end{align}
and which we, in analogy to the boundary conditions on the circle, refer to as the Neveu-Schwarz type. Other possible boundary conditions are
\begin{equation}
\lambda_{m,1}|_0 = \lambda_{m,1}|_L = 0 \hspace{1cm}\text{and}\hspace{1cm}
\partial_x\lambda_{m,2}|_0 = \partial_x\lambda_{m,2}|_L = 0
\end{equation}
with resulting standing-wave expansion
\begin{align}
\begin{cases}
\lambda_{m,1}(x) = \sqrt{\frac{2}{L}}\sum_{k>0} \hat{\lambda}_{m,1}(k)\sin\del{\frac{\pi k x}{L}}\\
\lambda_{m,2}(x) = \sqrt{\frac{2}{L}}\sum_{k>0} \hat{\lambda}_{m,2}(k)\cos\del{\frac{\pi k x}{L}} + \sqrt{\frac{1}{L}} \hat{\lambda}_{m,2}(0)
\end{cases}
\end{align}
to which we refer as the Ramon type. In both cases, the prefactors where chosen such that the Majorana modes obey their usual anti-commutation relations 

\begin{align}
	\{\hat{\lambda}(k)_{m,i}  ,  \hat{\lambda}(l)_{n,i}   \}= \delta_{m,n}\delta_{k,l} \delta_{i,j}.
\end{align}
To construct a Fock space we need to define normal fermionic modes. For $k \neq 0$ we can define $\hat{\phi}_m(k) = \hat{\lambda}_{m,1}(k) + \mathrm{i}\hat{\lambda}_{m,2}(k)$, which again transform under the fundamental (i.e.\ vector) representation of $\mathsf{SO}(2N)$. The $2N$ zero zero modes $\hat{\lambda}_{m,2}(0)$ are grouped into $N$ additional fermions $\alpha_{c}$ with $c \in 1,\ldots,N$. In terms of these operators  and after proper normalisation so that $\text{Tr}(e^{-2\pi H_E}) = 1$, the resulting entanglement Hamiltonian is given by

\begin{align}
	H^{(\text{NS})} = &\sum_{k=1}^{+\infty}\sum_{m=1}^{2N} \frac{\pi(k-1/2)}{L}\phi^\dagger_m(k)\phi_m(k)   +\frac{N}{\pi}\sum_{k=1}^{+\infty}\log\del{1+e^{-2\pi \frac{\pi(k-1/2)}{L}}   } \label{NS}
\end{align}
for the Neveu-Schwarz boundary conditions and 
\begin{align}
	H^{(\text{R})} =& \sum_{k=1}^{+\infty}\sum_{m=1}^{2N} \frac{\pi k}{L}\phi^\dagger_m(k)\phi_m(k) + 0\sum_{c=1}^{N} \alpha_c^\dagger\alpha_c  +\frac{N}{\pi} \sum_{k=1}^{+\infty}  \log\del{1+e^{-2\pi \frac{\pi k}{L}}}   + \frac{N}{2\pi}\log(2) \label{R}
\end{align}
for Ramon boundary conditions. Due to the zero modes, all eigenvalues of $H^{(R)}$ will be at least $2^N$ fold degenerate. In particular, the ground state will be an $\mathsf{SO}(2N)$ scalar in the Neveu-Schwarz case and a direct sum of the two fundamental spinor representations in the Ramon case. More generally, as higher excited states are obtained by acting with the vector operators $\phi_m^\dagger$ on those ground states, all eigenspaces of $H_{E}^{(\text{NS})}$ will transform as tensor representations, whereas all eigenspaces of $H_{E}^{(\text{R})}$ will transform according to spinor representations. It is thus straightforward to relate these two towers of eigenvalues to the entanglement spectrum obtained across cuts corresponding to virtual bonds with linear representations (for NS) and with projective representations (for R). 

\subsection{Numerics for $N=2$}

\begin{figure}[t!]
	\centering
	\includegraphics[width=0.6\columnwidth]{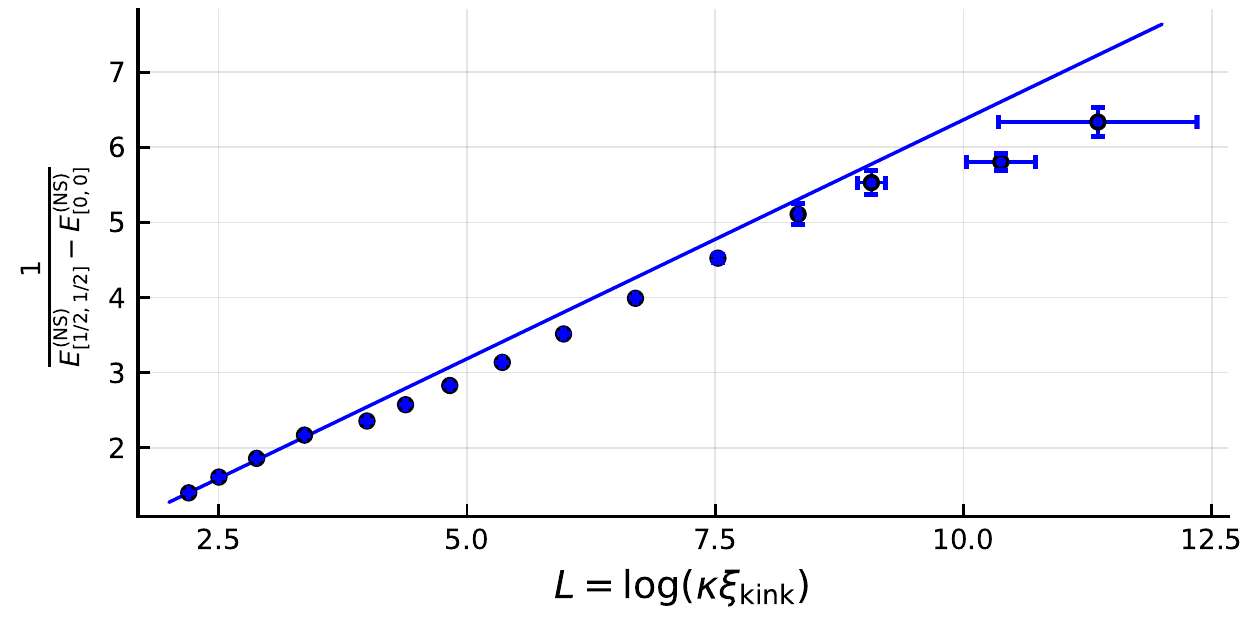}
	\caption{The inverse gap between the first two eigenvalues of the entanglement Hamiltonian in the trivial sector as a function of $L = \log(\kappa\xi_{\text{kink}})$. The error bars on the singular values are obtained in a similar fashion to those of the inverse correlation length (see Fig.~\ref{extrapol} ). The blue line is the CFT prediction $ \frac{2}{\pi}L$ and describes the extrapolated data well for $\kappa\approx2.83$. }
	\label{L}
\end{figure}

\begin{figure*}[t!]
	\begin{minipage}{.6\textwidth}
		\vspace*{1.1cm}
		\includegraphics[width=\textwidth, height=6cm]{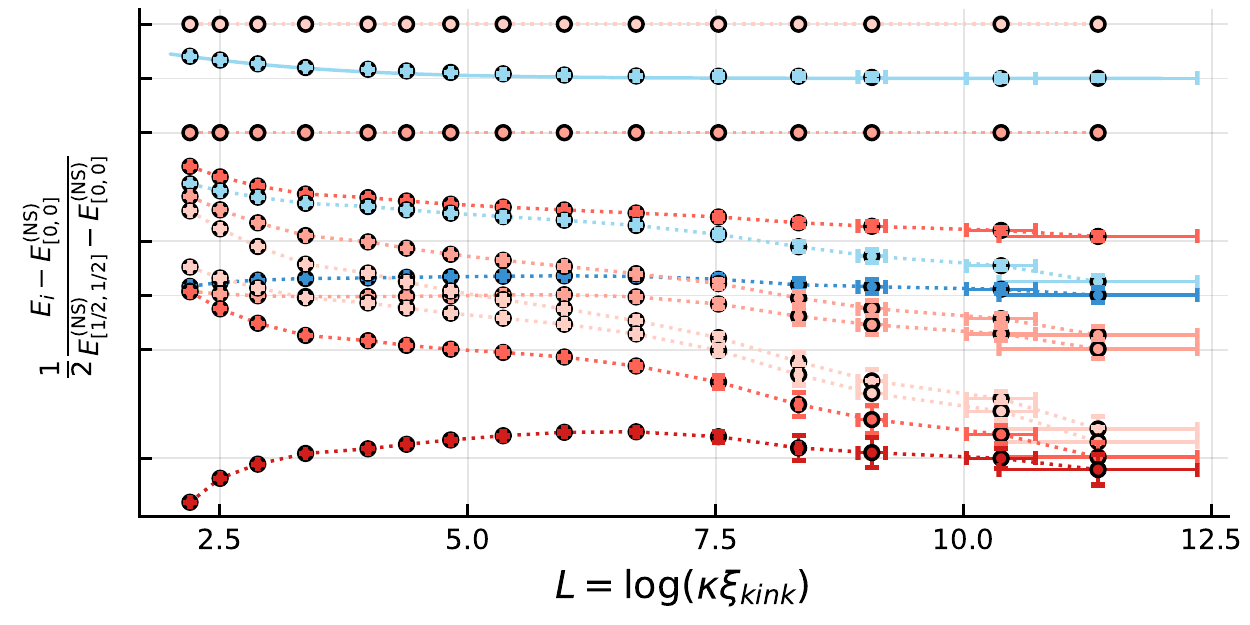}
	\end{minipage}
	\begin{minipage}{.38\textwidth}
		\includegraphics[width=\textwidth, height=5.5cm]{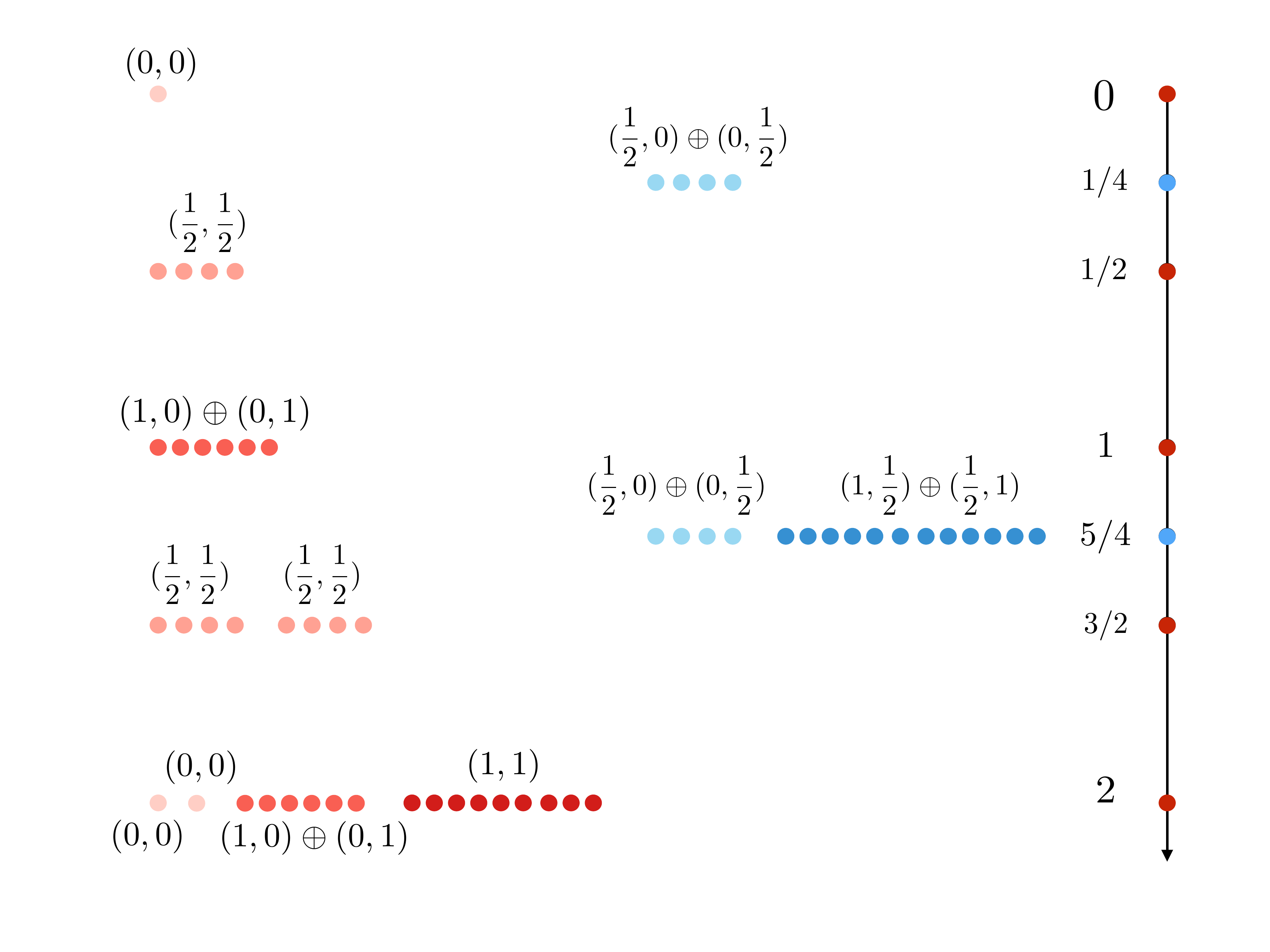}
	\end{minipage} 
	\caption{On the right we show the CFT prediction for the gaps in the entanglement spectrum, for $N=2$ and in the limit of large $L$. The vertical axis is rescaled by $\pi/L$. The left and right towers corresponds to NS and R boundary conditions, respectively. The plot in the left shows gaps in the entanglement spectrum, relative to the first gap in the NS sector, as extracted from the MPS representation of the ground state of our lattice model. Error bars are obtained in a similar fashion to those of the inverse correlation length (see Fig.~\ref{extrapol}). For sufficiently small couplings the ratios converge to those predicted in the right panel. The dotted lines are a guides for the eye. The full blue line is the prediction for the first gap in the Raymond sector from \ref{GS difference}. } 
	\label{towers}
\end{figure*}

For $N=2$, the lowest excited state of the entanglement Hamiltonian with NS boundary conditions, i.e. k=1 in Eq.~\ref{NS} is a $(1/2,1/2)$ quartet (the $\mathsf{SO}(4)$ vector representation) with gap : \begin{align}
	E^{(\text{NS})}_1 - E^{(\text{NS})}_0 =  \frac{\pi}{2L}\label{L_def}
\end{align}
Where $L = \log(\kappa \xi_{\text{kink}})$ is the typical length scale of the system after the comformal mapping. The free fit parameter $\kappa$ takes care of setting the UV scale. In Fig.~\ref{L} we show the inverse gap $\frac{1}{E_1^{(\text{NS})}-E_0^{(\text{NS})}} = 2\pi \del{\log(\lambda_1^{(\text{NS})}-\lambda_0^{(\text{NS})} )}^{-1}$ as a function of $L$, with $\lambda_i^{(\text{NS})}$ the $i$-th largest eigenvalue of $\hat{\rho}^{(\text{NS})}$, the reduced density matrix for a cut across the MPS bond with linear representations. The line corresponds to the prediction $\frac{2}{\pi}L$, and for $\kappa = \kappa_{\text{fit}} \approx 2.83$ the data coincides with this prediction.\\  
Filling the $k=1$ mode twice results in an energy $\frac{\pi}{L}$ sextet (the antisymmetric rank-2 tensor representation of $\mathsf{SO}(4)$). There are two possibilities to obtain energy $\frac{3}{2}\frac{\pi}{L}$, namely by filling the mode $k=2$ (a vector representation), or filling the $k=1$ mode with 3 particles (an antisymmetric rank-3 tensor representation, which is equivalent to the vector representation for $\mathsf{SO}(4)$). For the difference between the ground state energy in the two sectors we find
\begin{align}
E_0^{(\text{R})} &- E^{(\text{NS})}_0 = \frac{\log(2)}{\pi} + \frac{2}{\pi}\sum_{k=1}^{+\infty}\log\del{\frac{1+e^{-2\pi \frac{\pi k}{L}}}{1+e^{-2\pi \frac{\pi (k-1/2)}{L}}}} \label{GS difference}
\end{align}
which can be shown to converge to $\frac{\pi}{4L}$ for sufficiently large $L$ (or thus, exponentially large correlation lengths). Here, we have assumed that the same length parameter $L$ can be used for the two types of boundary conditions (NS or R). These eigenvalues, and a few more, for both $H^{(\text{NS})}$ and $H^{(\text{R})}$, relative to $\pi/L$, are depicted in the right panel of Fig.~\ref{towers}. In the left panel we show the corresponding ratios obtained trough extracted MPS data. The gaps  $-\frac{1}{2\pi}\log\del{{\lambda_i}/{\lambda^{(NS)}_0 }}$ are rescaled by twice the gap in the NS sector which according to Fig.~\ref{L} is approximatly $\pi/L$. As anticipated, towards the continuum limit (i.e. large $L$) all ratios converge to the predicted value. Note, however, that significant lattice effects for smaller $L$  are present, as the logarithmic mapping in the definition of the modular Hamiltonian makes it exponentially harder for the entanglement spectrum (as compared to \textit{e.g.}\ the excitation spectrum) to correspond with the continuum limit. In particular, additional degeneracies between different $\mathsf{SO}(N)$ representations which are predicted by the CFT result are not exactly reproduced away from the continuum limit. Finally, the predicted value for the gap between the groundstate energies in the Raymond and Neveu Schwarz sector (cfr. Eqs.~\ref{GS difference} and ~\ref{L_def}) is also shown; it matches well with the data even for smaller $L$ where this is not expected.

 \begin{figure}[t!]
	\centering
	\includegraphics[width=0.6\columnwidth]{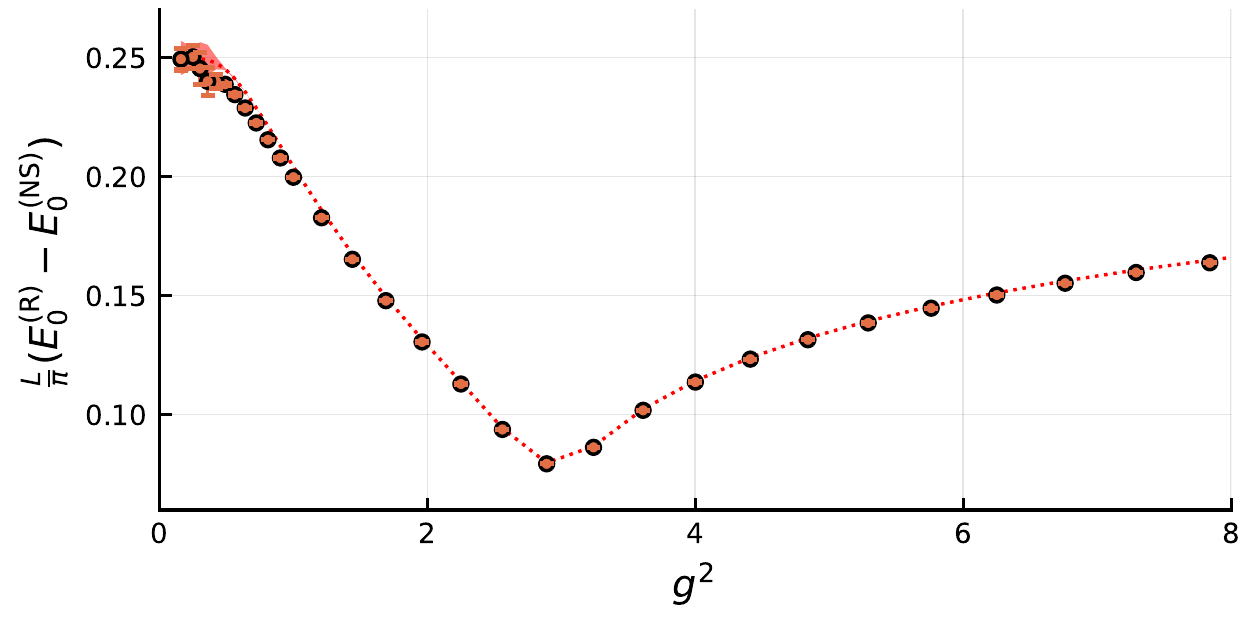}
	\caption{The gap between the two ground states of the modular Hamiltonian as a function of the squared coupling. The dotted line represents the CFT prediction from Eq.~\eqref{GS difference}, where $L$ was eliminated using Eq.~\eqref{L_def}, so that no free parameters remain. The error bars and shaded area correspond to uncertainties in the extrapolated entanglement eigenvalues.}
	\label{topfromtriv}
\end{figure}

To further highlight this last peculiar observation, we use the gap between the dominating entanglement eigenvalues in the trivial sector to predict the gap for the topological sector, by eliminating $L$ between Eq.~\eqref{GS difference} and Eq.~\eqref{L_def}. Fig.~\ref{topfromtriv} compares this prediction to the actual values of $E_0^{(\text{R})} - E^{(\text{NS})}_0 = -\frac{1}{2\pi}\log\del{{\lambda^{(R)}_0}/{\lambda^{(NS)}_0 }}$. We obtain excellent agreement, even for large couplings far away from the continuum limit, where the CFT prediction is no longer expected to hold. Indeed, beyond $g^2 \gtrsim 4$, the correlation length $\xi$ is less than a lattice site.

 \begin{figure}[t!]
	\centering
	\includegraphics[width=0.6\columnwidth]{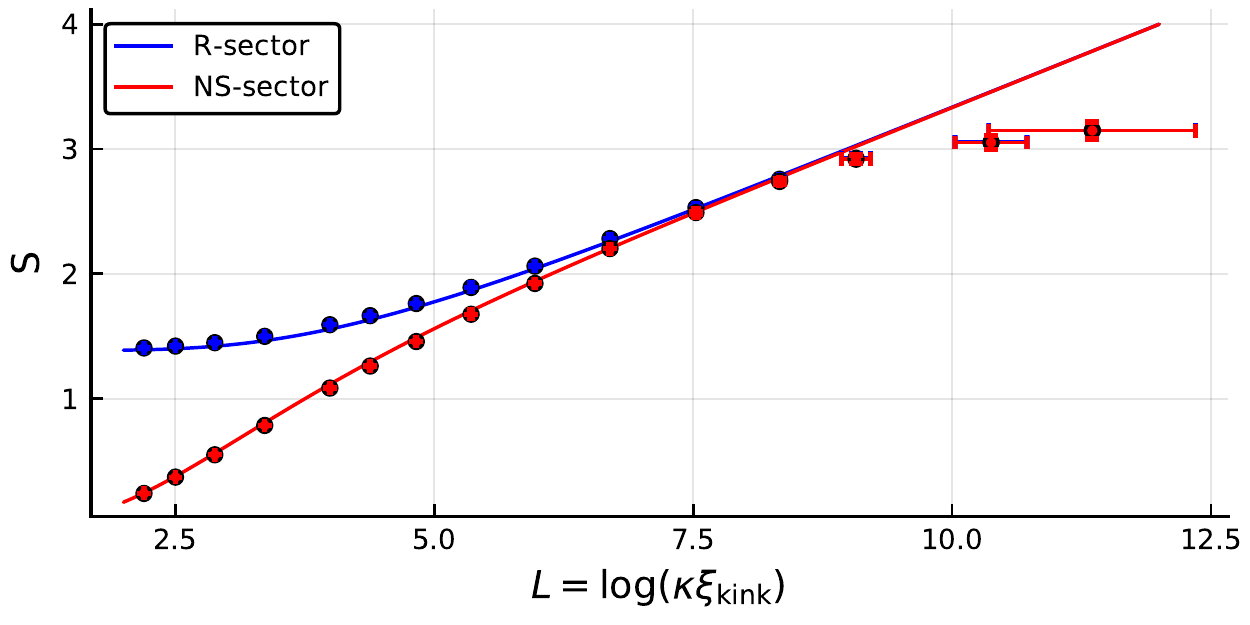}
	\caption{The bipartite entanglement entropy along the different cuts as a function of the logarithmic length scale $L$. The error bars are calculated in a similar fashion to those in Fig.~\ref{extrapol}. The curves represent the entropies corresponding to the entanglement spectra predicted from entanglement hamiltonians Eqs.~\ref{NS} and ~\ref{R}. We find good agreement except for very large $L$; this is due to ill converged MPS results, as was also the case in Fig.~\ref{relativity_check}.}.
	\label{entropy}
\end{figure}

Finally, Fig.~\ref{entropy} shows the scaling of the total bipartite entanglement entropy along trivial and topological cuts of the MPS as a function of the logarithmic length scale $L=\log(\kappa_{\text{fit}}\xi_{\text{kink}})$. The curves are the predicted entropies calculated from reduced density matrices corresponding to Eqs.~\ref{NS} and ~\ref{R}. Again there is good agreement apart from large $L$ where the MPS results are not sufficiently converged.

\section{Discussion and outlook}
\label{sec:ending}
We have constructed a lattice regulated version of the GN model that preserves the full $\mathsf{O}(2N)$ symmetry, and has a lattice remnant of the discrete chiral invariance. This prescription differs from the typical regularisation using Wilson fermions (i.e.\ the Gross-Neveu-Wilson model, which has recently received attention from the cold atoms community), and also takes a different prescription for the mass term as proposed in the original staggered formulation by Susskind. This prescription is well known in the context of the SSH model, but a Gross-Neveu type interaction resulting from it had, to the best of our knowledge, not been considered.

By studying this lattice model in the limit of large $N$ ---where mean field theory becomes exact--- as well as at $N=2$ using MPS simulations, we have established that its low energy behaviour replicates all the features (and in particular degeneracies) expected from the field theory. At the same time, we argued how the resulting lattice model lies at the first order phase transition between a trivial and topological insulator (according to symmetry class BDI in the ten-fold way), and much of the degeneracies in both the excitation and entanglement spectrum can be reinterpreted from that perspective. At the quantitative level, we observed that the non-perturbative behaviour of this marginally relevant interaction makes it especially challenging to accurately probe the continuum limit. The mass remains very small for a significant range of the coupling constant, and then shoots up quickly, so that the regime where MPS can probe the behaviour of the continuum limit is rather small. 

As the spectrum of massive particles becomes more interesting for larger values of $N$, it would be interesting to also study the model in this regime. However, our results on the entanglement structure indicate why using MPS simulations for larger values of $N$ is non-trivial. The entanglement structure, and in particular the entanglement entropy, is dominated by the UV CFT of the model, which is that of $N$ massless free fermions. We thus anticipate a linear scaling of the entanglement entropy in the number of fermion flavours, which translates to an exponential scaling in $N$ of the required MPS bond dimension, in order to obtain similarly accurate results. It is an interesting question whether exploiting the full $\mathsf{(S)O}(2N)$ symmetry of the model could help to overcome this exponential scaling. However, this first requires that the necessary representation data (Clebsch-Gordan coefficients and/or 6j-symbols) of $\mathsf{(S)O}(2N)$ are computed, as these are less readily available for general $N$.

Other potentially interesting directions of further research concern the phase diagram for finite values of the temperature and chemical potential, which are also within the scope of MPS simulations \cite{weir2010studying,takeda2015grassmann,banuls2015thermal,buyens2016hamiltonian,silvi2017finite}. There is active interest in the possible existence of an inhomogeneous phase at sufficiently large values of the chemical potential \cite{karsch1987gross,lenz2020inhomogeneous}. This interest is again spurred by the similarity of the GN model with QCD. Due to the sign problem, probing the QCD phase diagram with lattice Monte Carlo at moderate densities and with realistic values of the quark masses (the regime interesting for heavy ion experiments) is near impossible. While an inhomogeneous phase in the GN phase diagram would result in breaking of translation invariance, a continuous symmetry of the field theory, there might be arguments to believe that the Coleman-Mermin-Wagner theorem does not apply and the GN model could indeed exhibit such a phase. Coleman explicitly assumed relativistic invariance in his version of the theorem, which is broken by the chemical potential, whereas more general arguments against continuous symmetry breaking rely on the specific dispersion relation and the counting of the would-be Goldstone bosons that restore the symmetry, which is non-trivial when breaking spacetime symmetries. It would be interesting to study if MPS techniques can shed a new perspective on this question, though infinite MPS simulations would also need to choose a particular unit cell and would also struggle with incommensurate filling fractions.

A final extension, which we explore in a future publication \footnote{G.~Roose \textit{et al.}, \textit{in preparation}.}, is to apply the discretisation scheme presented in this paper to the chiral extension (with full continuous chiral symmetry) of the Gross-Neveu model. Preliminary results indicate that the resulting lattice model has emerging continuous chiral symmetry along a critical line in the phase diagram that corresponds to a deconfined quantum phase transition.

\acknowledgments{We acknowledge valuable discussions with Henri Verschelde, Erez Zohar , Bram Vanhecke, Maarten Van Damme and Daan Maertens. This work has received funding from the European Research Council (ERC) under the European Unions Horizon 2020 research and innovation programme (grant agreements No 715861 (ERQUAF) and 647905 (QUTE)), and from Research Foundation Flanders (FWO) via grant GOE1520N.}

\appendix
\section{Matching the lattice regularisation with $\overline{\text{MS}}$ dimensional regularisation}
\label{sec:appendix}
\begin{figure}
	\centering
	\includegraphics[width=0.6\columnwidth]{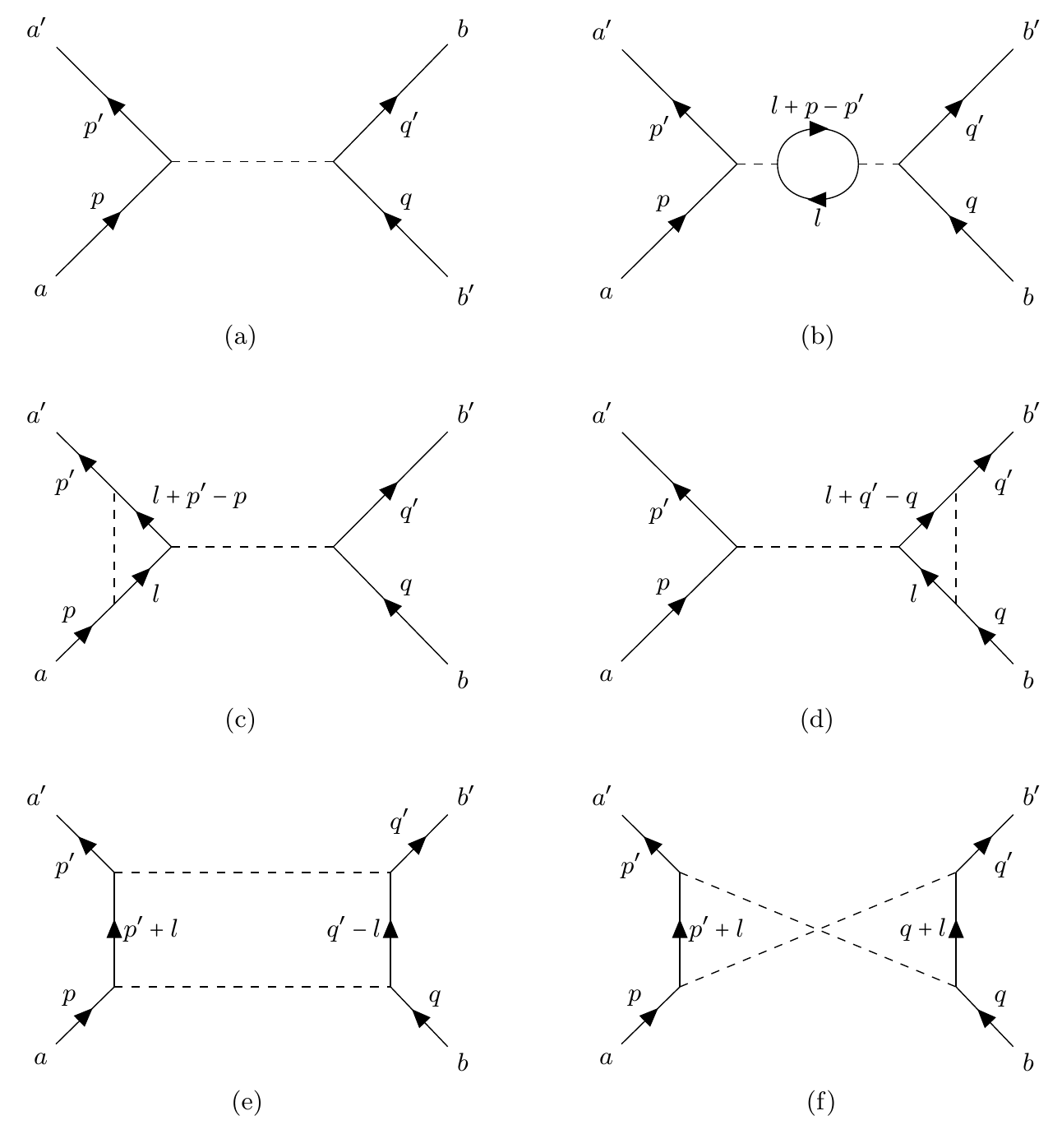}
	\caption{The Feynman diagrams contributing to the two fermion scattering $S$-matrix up to second order.}
	\label{scatter}
\end{figure}

The beautiful result Eq.~(\ref{eq:exactmassgap}) of Forgacs \textit{et al.}\ \cite{forgacs1991exact} gives the exact mass gap for the $\mathsf{O}(2N)$ Gross-Neveu model in terms of $\Lambda_{\overline{\text{MS}}}$, where the latter is given by Eq.~\eqref{RGmass}, but with coupling $g^2_{\overline{\text{MS}}}(\mu=1/a)$ of the $\overline{\text{MS}}$-scheme instead of our lattice coupling $g^2$. To relate  $\Lambda_{\overline{\text{MS}}}$ to our $\Lambda_{\text{lat}}$, one needs to match the dimensional regularisation scheme to our lattice regularisation scheme. In particular, we require the first coefficient $c_1$ in the expansion 
\begin{equation}
\frac{1}{g^2_{\overline{\text{MS}}}}=\frac{1}{g^2}+c_1+c_2 g^2+\ldots\,.
\end{equation}
The standard strategy to obtain such a matching is to compare results for a physical quantity, which by definition should be independent of the particular renormalisation scheme. We consider the two-fermion scattering $S$-matrix in the large energy/momentum regime, where perturbation theory is reliable, and the physics is well described by weakly interacting massless Dirac fermions.  Notice that for the lattice regularised version `large energy/momentum' $E$ means $\Lambda_{\text{lat}}\ll E \ll  1/a$, with the first inequality assuring the weakly interacting regime and the latter inequality assuring the QFT continuum regime.  In Fig.~\ref{scatter} we display the different Feynman diagrams that contribute up to one loop to this scattering process. Notice that, as in the original Gross-Neveu paper \cite{gross1974dynamical}, it is convenient to decompose the quartic term $\frac{g^2}{2}\del{\bar{\psi_a}\psi_a }^2$ in the QFT (\ref{GN:QFT}) (or $- \frac{g^2}{4}\Sigma_{n,n+1,n+2}^2 $ in the lattice Hamiltonian (\ref{eq:gnlattice})) by introducing a Hubbard-Stratanovich field $\sigma$ with trivial propagator $-i$ and interactions:  
\begin{align}
	 -g \sigma \bar{\psi_a}\psi_a \hspace{0.5cm}\text{ in the QFT and }\hspace{0.5cm} \frac{g}{\sqrt{2}}\sigma\Sigma_{n,n+1,n+2} \hspace{0.5cm}\text{ on the lattice }\label{eq:int}
\end{align}

For reference, we first briefly discuss the computation in the $\overline{\text{MS}}$ scheme. To get the Feynman rules one first needs the free-field propagator (see e.g. \cite{Peskin})
\begin{equation}
\bra{0}T(\psi_a(x^0,x^1)\overline{\psi}_b(y^0,y^1))\ket{0}=\int \frac{d^{2}p}{2\pi^2}\frac{i \slashed {p}}{p^2+i\epsilon} e^{-i p(x-y)}
\end{equation} using relativistic notation (with e.g. $p^2=p_0^2-p_1^2$, $x=(x^0,x^1)$).

The Feynman rules then read: 
\begin{eqnarray} 
\diagramw{2.5cm}{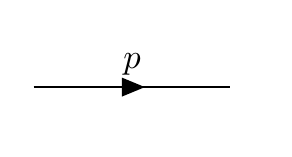} &=& \frac{i \slashed {p}}{p^2+i\epsilon} \nonumber\\
\diagramw{2.5cm}{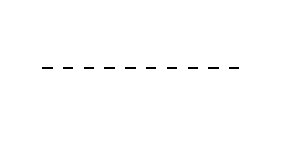}    &=& -i \nonumber\\
\diagramw{2.5cm}{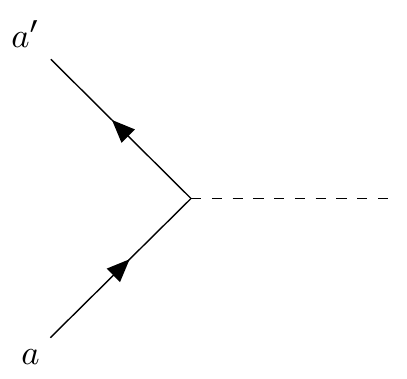}            &=& -i g \delta^a_{a\prime} \label{msbarfeynmanrules} 
\end{eqnarray} 

With the conventions of \cite{Peskin}, writing $S=1+i T$, the tree-level diagram (a) then gives for $iT$:
\bea (a)&=& ig^2\times\underbrace{(\overline{u}(p^{\prime})u(p)) (\overline{u}(q^{\prime})u(p))}_{K_1}  \times\underbrace{\delta^a_{a\prime} \delta^b_{b\prime} (2\pi)^2\delta^2(p+q-p^{\prime} -q^{\prime})}_{K_2} \eea
with the second factor $K_1$ on the first line arising from the projection on the particular fermion polarisations of the in- and out-modes that we consider, while the third factor $K_2$ (second line) arises from the colour conservation and momentum conservation. Notice that we do not consider the crossing diagrams for the outgoing legs, which gives terms in $iT$ proportional to a different colour pre-factor $\delta^a_{b\prime} \delta^b_{a\prime}$.

Using the standard machinery of dimensional regularisation and `Diracology', we then obtain for the second diagram (b) in the $\overline{\text{MS}}$ scheme:
\be (b)= - i \frac{N g^4}{2\pi} \log\frac{-(p-p^{\prime})^2}{\mu^2}\times K \ee
with $K=K_1\times K_2$ the same polarisation factor and colour/momentum conservation factor as for (a).
Furthermore for the diagrams (c) and (d) we find:
\be (c)+(d)= + i \frac{g^4}{2\pi} \log\frac{-(p-p^{\prime})^2}{\mu^2}\times K \ee
Finally one can verify that the individual (logarithmic) UV divergencies in diagrams (e) and (f) cancel out when summed together. (e)+(f) is therefore scheme independent and plays no role in the matching. Notice that these separate UV divergencies in (e) and (f) would require a marginal counter term $\propto (\overline{\psi}_a\gamma_\mu\psi_a)(\overline{\psi}_b\gamma^\mu\psi_b)$ which is prohibited by the full $\mathsf{O}(2N)$ symmetry. 

Collecting the different terms together we finally find up to order $g^4$  (neglecting scheme independent terms and terms $\propto \delta^a_{b\prime} \delta^b_{a\prime}$):
\be iT=ig^2(1-\frac{(N-1) g^2}{2\pi}\log\frac{-(p-p^{\prime})^2}{\mu^2})\times K \label{TMSbar}\ee
for the $S$-matrix in the $\overline{\text{MS}}$ scheme. 

Let us now turn to the computation of the same diagrams, but now with our lattice Hamiltonian (\ref{eq:gnlattice}). As a first ingredient we consider the free-field propagator. From the free Hamiltonian (\ref{eq:dirackinetic}) one easily shows:   
\bea   \bra{0} T( \phi_{m}(t)^\dagger \phi_{n}(u))\ket{0}=&\theta(t-u)\int^\pi_0 \frac{dk^1}{2\pi} e^{ik^1(m-n)-i\omega(k^1)(t-u)}\nonumber\\
-&\theta(u-t)\int^\pi_0 \frac{dk^1}{2\pi} e^{ik^1(n-m)-i\omega(k^1)(u-t)}\nonumber \eea
with $\omega(k^1)=2\sin|k^1|$, the particle energies (in lattice units) corresponding to the momenta $k^1$ (and we have omitted the colour indices). 
Using the Fourier representation of the step-function $\theta(t)$ we can rewrite the expression for the propagator above as:
\bea \int_{-\infty}^{+\infty}\!\! \frac{dk^0}{2\pi}\int^\pi_{-\pi}\!\! \frac{dk^1}{2\pi}\left(\frac{\theta(k^1)\, i}{k^0 -\omega(k^1)+i\epsilon} + \frac{\theta(-k^1)\, i}{k^0 +\omega(k^1)-i\epsilon}\right)
 e^{ik^1(m-n)-ik^0(t-u)} \label{eq:lattprop}\eea Notice that here we are not blocking the staggered sites, and the two Dirac spinor components now transpire in the two different branches $-\pi/2\leq k^1<\pi/2$ and $\pi/2\leq k^1<3\pi/2$ of the (angular) spatial momentum $k^1$ of our single component fermions. In particular, with the identification of the \emph{physical} (lattice-independent) momentum 
\be p^1=\frac{2k^1}{a}\label{physicalR}\ee  
we have $\phi_{2n+1} \approx \phi_{2n}$ for small momenta $|p^1|\ll 1/a$, corresponding to the Fourier transformed right-handed Dirac component $\psi_R$ in the Weyl representation, while the identification \be p^1=\frac{2k-2\pi}{a} \label{physicalL}\ee gives $\phi_{2n+1} \approx -\phi_{2n}$ for $|p^1|\ll 1/a$,  corresponding to the Fourier transformed left-handed Dirac component $\psi_L$. One can easily verify that these identifications give the correct positive energy $k^0>0$ poles (in lattice units) in the first term of our propagator (\ref{eq:lattprop}) $k^0\approx  a |p^1|$, both for the right-moving fermions ($p^1>0$) and left-moving fermions ($p^1<0$), while the second term of our propagator gives the proper poles for the anti-fermions. 

We are now ready to write down the Feyman rules for our Hamiltonian (\ref{eq:gnlattice}) with decomposed interaction term (\ref{eq:int}) :
\begin{eqnarray} 
\diagramw{2.5cm}{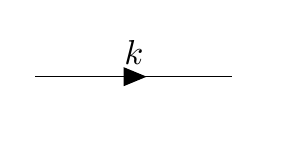} &=& \frac{i(k^0+2\sin k^1)}{{k^0}^2 -4\sin^2k^1+i\epsilon}  \nonumber\\
\diagramw{2.5cm}{Diagram_prop_scal.pdf} &=& -i\\
\diagramw{2.5cm}{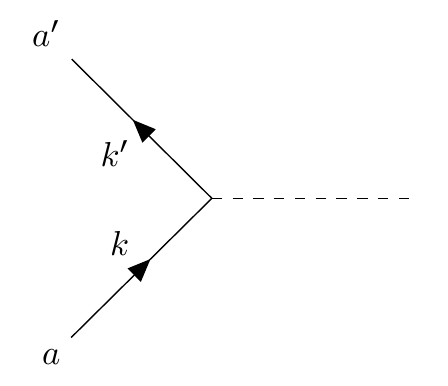} &=&  \frac{g}{\sqrt{2}}\delta^{a}_{a\prime} \left(\cos (k^{1\prime})-\cos (k^1)\right)\nonumber\,.
\end{eqnarray} 
Here for the expression of the fermion-propagator we summed the two terms in (\ref{eq:lattprop}). Also notice the extra momentum structure in the vertex vis-\'{a}-vis the vertex rule (\ref{msbarfeynmanrules}) in the $\overline{\text{MS}}$-scheme. For small physical momenta $p^1$ (see (\ref{physicalR}) and (\ref{physicalL})) this structure simply expresses the fact that the $\bar{\psi}\psi$ term couples the right-handed Dirac components ($k^1\approx 0$) to the left-handed components ($k^1\approx \pi$). 

For the tree-level diagram (a) in Fig.~\ref{scatter} we then find (replacing $p\rightarrow k$ and $q\rightarrow h$): 
\bea (a)&=& i g^2\times\underbrace{\left(\cos(k^{1\prime})-\cos(k^1)\right)\left(\cos(h^{1})-\cos(h^{1\prime})\right)}_{K_1}\nonumber\\
&&\hspace{3cm}\times\underbrace{\delta^a_{a\prime} \delta^b_{b\prime} (2\pi)^2\delta^2(k+h-k^\prime -h^\prime)/2}_{K_2}\,\,\eea
Taking into account the different normalisations (e.g. $\braket{k|k^\prime}=2\pi \delta(k-k^\prime)$ here, and $\braket{p|p^\prime}=4\pi E_p \delta(p-p^\prime)$ in the QFT computation) one can show that this reduces to the tree-level $\overline{\text{MS}}$ result (as it should) for small momenta $|p^1a|\ll 1$, either on the right branch from Eq.~\eqref{physicalR} or on the left branch of Eq.~\eqref{physicalL}. Notice that for a non-vanishing scattering in this continuum limit, we either need $k^1,h^{1\prime}\approx 0$ (right-movers) and $k^{1\prime },h^1\approx \pi$ (left-movers) or the other way around. Since for the remainder we are only interested in the continuum QFT limit, we can effectively set $K_1=4$ and anticipate that $k^1-k^{1 \prime}=\pi +\Delta^1$, with $\Delta^1\ll1$. 

For the loop diagram (b) we then find (with again $K=K_1\times K_2$):
\be (b)= -\frac{N g^4}{2}\times I_1\times K \ee
where $I_1$ is the loop-integral:
\bea I_1&=&\int \frac{d^2 l}{(2\pi)^2}\Bigg\{\left(\frac{(l^0+\Delta^0)-2\sin(l^1+\Delta^1)}{(l^0+\Delta^0)^2-4\sin^2(l^1+\Delta^1)+i\epsilon}\right)\nonumber\\
&&\quad \times \left(\frac{l^0+2\sin l^1}{{l^0}^2-4\sin^2 l^1+i\epsilon}\right)(\cos(l^1+\Delta^1)+\cos l^1)^2\Bigg\}\nonumber\eea
here $\Delta^0=k^0-k^{0\prime}$ and as we already mentioned $\Delta^1=k^1-k^{1\prime}-\pi$.
By closing the contour for the $l^0$ integration either in the upper or lower complex plane, and with a proper change of variables, we then arrive at the following expression for $I_1$:
\begin{align*}
	I_1=i \int^0_{-\pi}\! \frac{dl^1}{2\pi}\Bigg\{\left(\frac{-\Delta^0+2\sin l^1-2\sin(l^1-\Delta^1)}{(\Delta^0-2\sin l^1)^2-4\sin^2(l^1-\Delta^1)}\right)\times\left(\cos l^1+\cos(l^1-\Delta^1)\right)^2\nonumber\\
	+\left(\frac{\Delta^0+2\sin l^1-2\sin(l^1+\Delta^1)}{(\Delta^0+2\sin l^1)^2-4\sin^2(l^1+\Delta^1)}\right)\times\left(\cos l^1+\cos(l^1+\Delta^1)\right)^2\Bigg\}
\end{align*}
With some effort one can then finally extract the continuum limit $|\Delta^\mu|\ll 1$ of this integral, by isolating the logarithmic divergencies $\Delta^\mu\rightarrow 0$ around $l^1=0$ and $l^1=-\pi$, arriving at the leading behaviour: 
\begin{align*}
	I_1= \frac{i}{2\pi}\left(4-12\log 2 + 2\log\left(-({\Delta^0}^2-4{\Delta^1}^2)\right) +\mathcal{O}({\Delta^\mu}^2)\right)
\end{align*}
Notice that only this leading behaviour corresponds to the Gross-Neveu QFT continuum limit, the higher order power corrections are specific to the lattice regularisation, in QFT speak they correspond to irrelevant perturbations of the Gross-Neveu QFT.

Moving over to the loop diagram (c), we obtain: 
\be (c)= -\frac{g^4}{4}\times I_2\times K\ee
with now the loop-integral $I_2$ reading (for the case $k^{1\prime}, h^{1}\approx 0$):
\begin{align*}
	I_2=-\int \frac{d^2 l}{(2\pi)^2}\Bigg\{\left(\frac{(l^0-\Delta^0)-2\sin(l^1-\Delta^1)}{(l^0-\Delta^0)^2-4\sin^2(l^1-\Delta^1)+i\epsilon}\right)\times \left(\frac{l^0+2\sin l^1}{{l^0}^2-4\sin^2 l^1+i\epsilon}\right)\\ \hspace{2cm}\times(\cos(k^{1\prime}+\cos(l^1-\Delta^1))\times (\cos l^1 +\cos( \Delta^1+k^{\prime}))\times (\cos l^1 +\cos(l^1-\Delta^1) )\Bigg\}
\end{align*}
Proceeding in a completely similar fashion as for the computation of $I_1$, we eventually find the leading continuum behaviour:
\be I_2= \frac{i}{2\pi}\left(12\log2-4-2\log\left(-({\Delta^0}^2-4{\Delta^1}^2)\right)\right)\ee It is easy to show that we get the same term from the diagram (d), while as we explained above we can forget about the (e) and (f) diagrams for the matching as these diagrams will be finite and universal for a manifest $\mathsf{O}(2N)$ symmetric regularisation scheme like ours.
Collecting all the relevant diagrams we then find for the QFT $S$-matrix in our lattice regularisation:
\be iT= ig^2(1-\frac{(N-1) g^2}{2\pi}(\log(\frac{-(p-p^{\prime})^2}{\mu^2})+2-6\log2))\times K\label{Tlatt} \ee
where $\mu=1/a$  and we have identified (see (\ref{physicalR}) and (\ref{physicalL})) the physical momenta and energies:
\be {\Delta^0}^2-4{\Delta^1}^2=\left((p^0-p^{0'})^2-(p^1-p^{1'})^2\right)\times a^2 \ee  

Finally, we are ready to match the two schemes. By comparing the lattice result in Eq.~\eqref{Tlatt} with the $\overline{\text{MS}}$ result in Eq.~\eqref{TMSbar}, and
demanding $iT=iT$ we immediately find:
\be \frac{1}{g^2_{\overline{\text{MS}}}}=\frac{1}{g^2}-\frac{N-1}{2\pi}(6\log 2 -2)+\ldots \ee
and if we plug this result in the definition (\ref{RGmass}) for $\Lambda$, we obtain:
\be \Lambda_{\overline{\text{MS}}}=\frac{8}{e}\Lambda_{\text{lat}}\ee
by which we have shown explicitly that the matching result based on the large $N$ mean-field computation of section \ref{sec:meanfield}, generalises to any finite $N$, in particular to $N=2$.  

\bibliography{library.bib}

\end{document}